\documentclass[twocolumn,preprintnumbers,floats,floatfix,nofootinbib,superscriptaddress]{revtex4-1}
\usepackage[applemac]{inputenc}
\usepackage{amsmath,amssymb}
\usepackage{graphicx}
\usepackage{color}
\begin{document}

\title{Multidomain Galerkin-Collocation method: spherical collapse of scalar fields II}

\author{M. A. Alcoforado}
\email{malcoforado@hotmail.com}
\author{R. F. Aranha}
\email{rafael.aranha@uerj.br}
\affiliation{Departamento de F\'{\i}sica Te\'orica - Instituto de F\'{\i}sica
A. D. Tavares, Universidade do Estado do Rio de Janeiro, R. S\~ao Francisco Xavier, 524. Rio de Janeiro, RJ, 20550-013, Brazil}
\author{W. O. Barreto}
\email{wobarreto@gmail.com}
\affiliation{Departamento de F\'{\i}sica Te\'orica - Instituto de F\'{\i}sica
	A. D. Tavares, Universidade do Estado do Rio de Janeiro, R. S\~ao Francisco Xavier, 524. Rio de Janeiro, RJ, 20550-013, Brazil}
\affiliation{Centro de F\'{\i}sica Fundamental, Universidad de Los Andes, M\'erida 5101,  Venezuela}
\author{H. P. de Oliveira}
\email{henrique.oliveira@uerj.br}
\affiliation{Departamento de F\'{\i}sica Te\'orica - Instituto de F\'{\i}sica
	A. D. Tavares, Universidade do Estado do Rio de Janeiro, R. S\~ao Francisco Xavier, 524. Rio de Janeiro, RJ, 20550-013, Brazil}


\date{\today}

\begin{abstract}
We follow the strategy initiated in Ref. \cite{alcoforado} and proceed with the implementation of the Galerkin-Collocation domain decomposition (GCDD) applied to the dynamics of a spherical self-gravitating scalar field with the field equation in the Cauchy formulation. We have adopted the areal slicing gauge. We have presented a detailed implementation for an arbitrary number of subdomains and adopted the simplest form of the transmission conditions. Further, by an appropriate choice of the basis functions in the inner subdomain, we eliminated exactly the $1/r$ terms near the origin present in the field equations. The code is validated using two error measures: the conservation of the ADM mass and the Hamiltonian constraint that must be satisfied during the spacetime dynamics. In general, both error measures converge exponentially in all subdomains. As a useful illustration of placing more subdomains near the strong-field region, meaning an efficient concentrating of collocation points near the origin, we exhibited the formation of an apparent horizon even though the numerical integration diverges. 
\end{abstract}

\maketitle

\section{Introduction}
The present work is a continuation of the systematic development of the Galerkin-Collocation domains decomposition (GCDD) technique applied to problems of interest in numerical relativity. In Ref. \cite{alcoforado}, we have implemented an efficient algorithm to solve the field equations in the characteristic formulation describing the dynamics of a scalar field in spherically symmetric spacetimes, where we have included more than two subdomains. Here, our goal is to consider the same problem with the field equations in the Cauchy formulation.  We have already implemented GCDD codes successfully in the following cases: (i) determination of the initial data for single \cite{oliveira_14} and binary black holes \cite{barreto_18,barreto_18_2}, (ii) the dynamics of cylindrical gravitational waves \cite{barreto_19}, and (iii) critical collapse in the new characteristic scheme \cite{crespo_19}. 

{We also mention that this work also precedes studies in the near future which use the methods presented here.} {The context will be the BSSN formalism written in curvilinear coordinates \cite{nok87}--\cite{brown} which will be of great use in the analysis of more complex physical systems. As examples, we can highlight the collapse of scalar fields with general potentials, non-spherical gravitational collapse, collapse with rotation, in addition to the inclusion of more realistic matter. These cases leave spherical symmetry and require greater computational complexity (see Ref. \cite{deppe}, and references therein, as an example for critical collapse).  With that, this initial study proves to be crucial. It is also important to say that we intend to use the computational tools described in this work and previous articles to establish a general numerical relativity code via spectral methods. Our focus here is the computational method. } 

The domain decomposition or multidomain method consists of dividing the spatial domain into two or more subdomains and appropriate transmission conditions to connect the solutions in each subdomain. In the realm of spectral methods, domain decomposition methods were first established for fluid mechanics problems in the late 1970s. We indicate Canuto et al. \cite{canuto_88} (see also Ref. \cite{canuto_new}) for a helpful and concise presentation of domain decomposition methods. Orszag \cite{orszag_80} introduced the spectral domain decomposition method for elliptic problems; Kopriva \cite{kopriva_86,kopriva_89} considered the spectral multidomain technique for hyperbolic problems. In this case, we remark that there is no unique way of matching the solutions in contiguous subdomains \cite{kopriva_86,faccioli_96}.

The first application of the spectral domain decomposition method in Numerical Relativity was to determine the stationary configurations \cite{bona} and the initial data problem \cite{pfeifer,ansorg}. For the time-dependent systems, the spectral domain decomposition was implemented within the SpEC \cite{spec} and LORENE \cite{lorene} codes to deal with the gravitational collapse, the dynamics of stars, and the evolution of single \cite{kidder_00} and binary black holes \cite{szilagyi_09}. In particular, the use of discontinuous Galerkin methods {\cite{kidder1,kidder2,hw08}} has been implemented to describe the neutron star evolution successfully.  A more detailed approach for the multidomain spectral codes is found in Refs. \cite{Hemberger_13} and in the SXS collaboration \cite{sxs_col}. 

{We present an alternative numerical procedure within the domain decomposition spectral methods, which is currently in continuous improvement and development. The first new feature we have used systematically is the compactification of the spatial domain using the algebraic map \cite{boyd}, and in the sequence splitting it into several subdomains (cf. Fig. 1). Consequently, we introduce the map parameter, $L_0$, that controls the subdomains' locations and the distribution of collocation points. As we will show in the sequence, the convenient choice of the map parameter and the number of subdomains is essential for greater code accuracy and convergence. Another distinct feature worth mentioning is the choice of basis functions expressed as a combination of the Chebyshev polynomials to satisfy the boundary conditions. In particular, we have handled exactly the $1/r$ near the origin.} 

The sequence of presentation is as follows. In the second Section, we establish the basic equations. We describe the domain-decomposition strategy in Section 3. As a typical feature of the adopted numerical strategy, we have introduced a compactified intermediary computational domain before separating it into several subdomains. We define the basis functions in each subdomain after establishing the corresponding rational Chebyshev polynomials. In Section 4, we proceed with the numerical tests to validate the multidomain code. We have considered two robust indicators from which we defined the error measures: the conservation of the ADM mass and the Hamiltonian constraint verification. As the last test, we have determined the formation of the apparent horizon and its mass, considering we have adopted a coordinate system that the integration diverges when the apparent horizon forms. In the last Section, we summarize the results and indicate the next steps of the present investigation line.

\section{The field equations}%

We present briefly the main equations and the basic aspects of the spherical collapse of a scalar field in the Cauchy scheme of the field equations. We adopt the following line element:
\begin{equation}
ds^2=-\mathrm{e}^{2\alpha}\,dt^2 + \mathrm{e}^{2\mu}\, dr^2 + r^2(d \theta^2 + \sin^2 \theta d\varphi^2), \label{eq1}
\end{equation}

\noindent with $\alpha=\alpha(t,r)$ and $\mu=\mu(t,r)$ determing to the lapse function and the radial metric function, respectively. Assuming polar slicing and an areal radial coordinate, and considering a scalar field with potential $V(\phi)$, the relevant field equations are 
\begin{eqnarray}
&\alpha_{,r}-\mu_{,r}+\displaystyle{\frac{1}{r}}(1-\mathrm{e}^{2\mu})+r\mathrm{e}^{2\mu}V(\phi) = 0,& \label{eq2}\\
\nonumber \\
&\mu_{,t}=\displaystyle{\frac{r}{2}}\,\mathrm{e}^{\alpha-\mu}\Pi\phi_{,r},& \label{eq3}\\
\nonumber \\
&\mathcal{H}=\mu_{,r}+\displaystyle{\frac{1}{2r}}(\mathrm{e}^{2\mu}-1)-\frac{r}{4}(\Pi^2 + \phi_{,r}^2)-\displaystyle{\frac{r}{2}}\mathrm{e}^{2\mu}V(\phi)=0.& \nonumber \\
\label{eq4}
\end{eqnarray}


\noindent Eqs. (\ref{eq3}) and (\ref{eq4}) are the momentum and the Hamiltonian constraints, respectively. We solve Eq. (\ref{eq2}) to update the function $\alpha(t,r)$ in each time level or hypersurface slice, and we use Eq. (\ref{eq3}) to evolve the metric function $\mu(t,r)$ instead of solving the Hamiltonian constraint at each slice (see, for instance \cite{alcubierre_potential, choptuik,shapiro_teukolsky_80}). 

The Klein-Gordon equation
\begin{eqnarray}
\Box \phi = \frac{d V}{d \phi} \label{eq5},
\end{eqnarray}

\noindent can be split into two first-order equations: 
\begin{eqnarray}
&\phi_{,t}=\mathrm{e}^{\alpha-\mu} \Pi, & \label{eq6}\\
\nonumber \\
&\Pi_{,t}=\displaystyle{\frac{1}{r^2}\left(\mathrm{e}^{\alpha-\mu} r^2\phi_{,r}\right)_{,r}-\mathrm{e}^{\alpha+\mu}\frac{dV}{d\phi}},& \label{eq7}
\end{eqnarray}

\noindent with Eq. (\ref{eq6}) defining the function $\Pi(t,r)$.

We integrate the field equations (\ref{eq2}),\,(\ref{eq3}),\,(\ref{eq6}) and (\ref{eq7}) starting with the initial data $\phi(t_0,r)$ and $\Pi(t_0,r)$. We determine the initial function $\mu(t_0,r)$ after solving the Hamiltonian constraint (\ref{eq4}). Then, from Eq. (\ref{eq2}) we obtain $\alpha(t_0,r)$, and from Eqs. (\ref{eq3}),\,(\ref{eq6}) and (\ref{eq7}) we determine the initial functions $\mu_{,t}(t_0,r)$, $\phi_{,t}(t_0,r)$ and $\Pi_{,t}(t_0,r)$. With these quantities we fix the functions $\mu,\phi, \Pi$ at the next time level, and the whole process repeats providing the evolution of the spacetime. We remark that we use the Hamiltonian constraint only to obtain the initial distribution $\mu(t_0,r)$. We further validate the code verifying the Hamiltonian constraint at each slice.

A relevant quantity is the mass function defined by: 
\begin{equation}
1-\frac{2 m}{r} = g^{\mu\nu} r_{,\mu} r_{,\nu} = \mathrm{e}^{-2\mu}, \label{eq8}
\end{equation}
	
\noindent that is, $m(t,r)$ measures the mass inside an sphere of radius $r$ at the intant $t$. It can be shown that an apparent horizon forms when $\mu \rightarrow \infty$. Related to the mass function, we have the ADM mass
\begin{equation}
M_{ADM}(t)=\lim_{r \rightarrow \infty} m(t,r). \label{eq9}
\end{equation}
	
\noindent Contrary to the Bondi mass, the ADM mass is conserved \cite{ADM,misner,MTW}. 
	
The boundary and coordinate conditions are a essential part to guarantee the regularity of the spacetime at the origin and the asymptotic flatness.
Then, near the origin, we have
\begin{eqnarray}
\alpha(t,r)&=&\mathcal{O}(r^2), \label{eq10}\\
\nonumber \\
\mu(t,r)&=&\mathcal{O}(r^2), \label{eq11}\\
\nonumber \\
\phi(t,r)&=&\phi_0(t)+{\cal{O}}(r^2), \label{eq12}\\
\nonumber \\
\Pi(t,r)&=&\Pi_0(t)+{\cal{O}}(r^2). \label{eq13}
\end{eqnarray}
	

\noindent {where $\phi_0(t)=\phi(t,r=0)$ and $\Pi_0(t) = \Pi(t, r=0)$}. We {point} out that the cancellation of the $1/r$ terms present in the field equations requires fulfilling the above conditions. As we are going to show, with the Galerkin-Collocation method, these conditions are satisfied exactly. {This is so because we construct a set of basis functions which satisfy the above asymptotic conditions.}

The asymptotic-flatness conditions are 
{
\begin{eqnarray}
\alpha(t,r)&=&\alpha_0(t)+\frac{\alpha_{-1}(t)}{r}+{\cal{O}}(r^{-2}), \label{eq14}\\
\nonumber \\
\mu(t,r)&=&\frac{M_{ADM}}{r}+{\cal{O}}(r^{-2}), \label{eq15}\\
\nonumber\\
\phi(t,r)&=&{\tilde\phi}_0+\frac{{\tilde\phi}_{-1}(t)}{r}+{\cal{O}}(r^{-2}), \label{eq16}\\
\nonumber\\
\Pi(t,r)&=&\frac{\Pi_{-2}(t)}{r^2}+{\cal{O}}(r^{-3}). \label{eq17}
\end{eqnarray}}
	
\noindent {for suitable functions $\alpha_0(t)$, $\alpha_{-1}(t)$, $\tilde{\phi}_0(t)$, $\tilde\phi_{-1}(t)$, and $\Pi_{-2}(t)$. We remark that for a massless scalar field we set $\tilde{\phi}_0=0$, but depending on the potential we may have $\tilde{\phi}_0 \neq 0$ and/or $\tilde{\phi}_{-1}=0$.} Also, from the asymptotic expansion of the function $\mu(t,r)$, we can read off the ADM mass.

The mass function satisfies the equation
\begin{eqnarray}
\frac{\partial m}{\partial r} = \frac{r^2}{4}\mathrm{e}^{-2 \mu}(\Pi^2+\phi_{,r}^2) +\frac{r^2}{2} V(\phi), \label{eq18}
\end{eqnarray}

\noindent that is obtained after differentiating Eq. (8) and taking into account the Hamiltonian constraint (\ref{eq4}). We calculate the ADM mass integrationg the above expression yielding
\begin{eqnarray}
M_{ADM} = \int_{0}^\infty\, \left[\frac{r^2}{4}\mathrm{e}^{-2 \mu}(\Pi^2+\phi_{,r}^2) +\frac{r^2}{2} V(\phi)\right] dr. \nonumber \\ \label{eq19}
\end{eqnarray}

\noindent We point out that the ADM mass calculation is performed more accurately using the above integral {if we compare 
with the asymptotic expression for the ADM mass given by Eq. (\ref{eq9}).}

\section{The multidomain Galerkin-Collocation method}\label{Section_III}

We describe the numerical procedure based on the Galerkin-Collocation method in multiple nonoverlapping subdomains applied to a self-gravitating scalar field described by Eqs. (\ref{eq2}),\,(\ref{eq3}),\,(\ref{eq6}) and  (\ref{eq7}). We follow the same line of the presentation we have done in Ref. \cite{alcoforado} for the case of the characteristic scheme.


\subsection{Spectral approximations}

We divided the physical domain $\mathcal{D}: 0\leq r < \infty$ into $n$ subdomains $\mathcal{D}_1: 0\leq r \leq r_{(1)}$,.., $\mathcal{D}_{l}: r_{(l-1)} \leq r \leq r_{(l)}$,.., $\mathcal{D}_n: r_{(n-1)} \leq r <\infty$, with $r_0=0$, and $r_{(1)},r_{(2)},..,r_{(n-1)}$ representing the interface between contiguous subdomains.

\begin{figure*}[htb]
\includegraphics[scale=0.5]{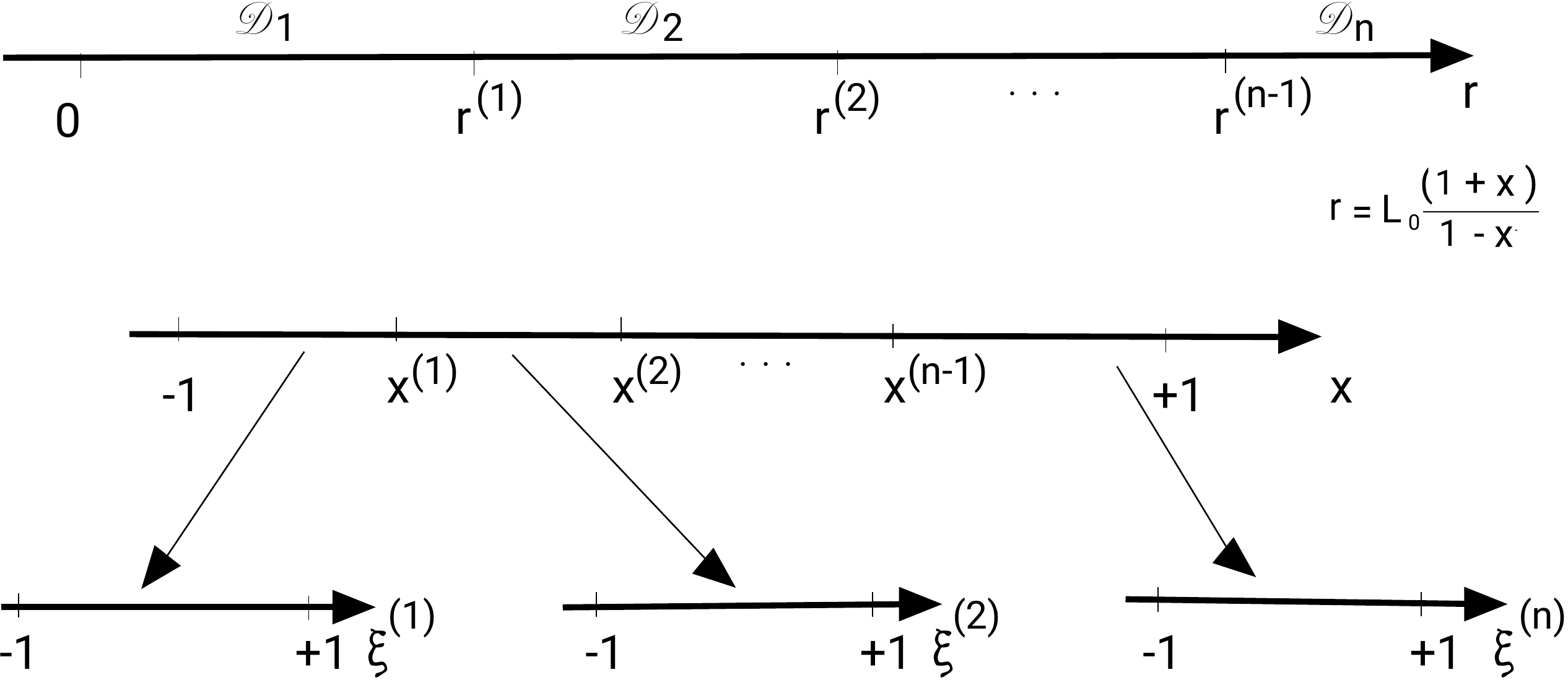}
	\caption{Basic scheme showing the subdomains $\mathcal{D}_1,\mathcal{D}_2,..,\mathcal{D}_n$ covered by $-1 \leq \xi^{(j)} \leq 1$, $k=1,2,..,n$, and the intermediate computational subdomain $-1 \leq x \leq 1$.}
\end{figure*}

In each subdomain, we establish the spectral approximations of the metric functions $\alpha(t,r),\,\mu(t,r)$ and the scalar field functions $\phi(r,t),\,\Pi(t,r)$:
\begin{eqnarray}
\alpha^{(l)}(t,r)&=&\sum^{N_\alpha^{(l)}}_{k=0}\hat{\alpha}^{(l)}_k(t)\Phi^{(l)}_k(r), \label{eq20}\\
\nonumber \\
\mu^{(l)}(t,r)&=&\sum^{N_\mu^{(l)}}_{k=0}\hat{\mu}^{(l)}_k(t)\Xi ^{(l)}_k(r), \label{eq21}\\
\nonumber \\
\phi^{(l)}(t,r)&=&\sum^{N_\phi^{(l)}}_{k=0}\hat{\phi}^{(l)}_k(t)\psi^{(l)}_k(r), \label{eq22}\\
\nonumber \\
\Pi^{(l)}(t,r)&=&\sum^{N_\Pi^{(l)}}_{k=0}\hat{\Pi}^{(l)}_k(t)\chi^{(l)}_k(r), \label{eq23} 
\end{eqnarray}


\noindent where $l=1,2,..,n$ indicates the specific subdomain; $\{\hat{\alpha}^{(l)}_k(t),\,\hat{\mu}^{(l)}_k(t),\,\hat{\phi}^{(l)}_k(t),\,\hat{\Pi}^{(l)}_k(t)\}$  are the modes or the unknown coefficients and $\{\Phi^{(l)}_k(r),\,\Xi^{(l)}_k(r),\psi^{(l)}_k(r),\,\chi^{(l)}_k(r)\}$ are  basis functions. We fix the number of modes in each subdomain by the truncation orders $N_\alpha^{(l)},N_\mu^{(l)},N_\phi^{(l)},N_\Pi^{(l)},\,l=1,2,..,n$. According with the Galerkin method, the basis functions belonging to the first and last subdomains must satisfy the boundary conditions (\ref{eq10}) - (\ref{eq13}) and (\ref{eq14}) - (\ref{eq17}), respectively. The basis functions in the intermediate subdomains $l=2,3,..,n-1$ are identified as the rational Chebyshev polynomials \cite{boyd} defined in these subdomain as we are going to define in the sequence.


\subsection{Rational Chebyshev polynomials}

We define the rational Chebyshev polynomials in each subdomain using Fig. 1 to indicate the mappings we have adopted to connect the physical and the computational subdomains. First, the domain $\mathcal{D}: 0 \leq r < \infty$ is mapped into the interval $-1 \leq x \leq 1$ using the algebraic map \cite{boyd}
\begin{equation}
r = L_0 \frac{(1+x)}{1-x}, \label{eq24}
\end{equation}

\noindent where $L_0$ is the map parameter. $x^{(1)},x^{(2)},..,x^{(n-1)}$ are the interface between contiguous subdomains as shown in Fig. 1. Second, we introduce linear transformations to define the computational subdomains parameterized by $-1 \leq \xi^{(l)} \leq 1$, $l=1,2,..,n$:
\begin{eqnarray}
x(\xi^{(l)})=\frac{1}{2}\,\left[\left(x^{(l)}-x^{(l-1)}\right)\xi^{(l)}+x^{(l)}+x^{(l-1)}\right], \label{eq25}
\end{eqnarray} 

\noindent with $l= 1...,n$ where $x^{(0)}=-1$ and $x^{(n)}=1$. These computational subdomains are the loci of the collocation points that are mapped back to $r_k$ in the physical domain $\mathcal{D}$.

The rational Chebyshev polynomials are defined in each subdomains as
\begin{eqnarray}
TL_k^{(l)} = T_k\left(\xi^{(l)}=\frac{a^{(l)}r+b^{(l)}}{(r+L_0)}\right), \label{eq26}
\end{eqnarray}

\noindent where $\xi^{(l)}$ is a combination of Eqs. (\ref{eq24}) and (\ref{eq25}) yielding  
\begin{eqnarray}
a^{(l)}&=&\frac{2L_0+r^{(l)}+r^{(l-1)}}{r^{(l)}-r^{(l-1)}}, \label{eq27}\\
\nonumber \\ 
b^{(l)}&=&-\frac{2r^{(l)}r^{{(l-1)}}+L_0\left(r^{(l)}+r^{(l-1)}\right)}{r^{(l)}-r^{(l-1)}}. \label{eq28}
\end{eqnarray}

\noindent $T_k(\xi)$ is the Chebyshev polynomial of kth order, $r^{(l-1)} \leq r \leq r^{(l)}$ corresponds to $-1 \leq \xi^{(l)} \leq 1$ for all $l=1,2,..,n$ with $r^{(0)}=0$ and $r^{(n)}$ is located at the infinity.


\subsection{Basis functions}

We can now define the basis functions for the spectral approximations (\ref{eq20}) - (\ref{eq23}). The rational Chebyshev polynomials (\ref{eq26}) are the basis functions for all spectral approximations in the subdomains $\mathcal{D}_l, l=2,3,..,n-1$. For the subdomains $\mathcal{D}_1$ and $\mathcal{D}_n$ the basis functions must satisfy the conditions (\ref{eq10}) - (\ref{eq13}) and (\ref{eq14}) - (\ref{eq17}), respectively.

The basis for the functions $\alpha(t,r)$ and $\mu(t,r)$ in the first subdomain $\mathcal{D}_1$ are
\begin{eqnarray}
\Phi_k^{(1)}(r) &=& \Xi_k^{(1)}(r) = \frac{(1+2k)}{4(3+2k)}TL_{k+2}^{(1)}(r)  \nonumber \\
&&+ \frac{(1+k)}{3+2k}TL_{k+1}^{(1)}(r)+\frac{1}{4}TL_{k}^{(1)}(r). \label{eq29}
\end{eqnarray}

\noindent As a consequence, we have $\Phi_k^{(1)}(r)=\mathcal{O}(r^2)$. Concerning the scalar field basis functions, we have adopted a distinct approach by using the Tau method \cite{boyd}. To reproduce the conditions (\ref{eq12}) and (\ref{eq13}), we assume that $\psi^{(1)}_k(r)=\chi^{(1)}_k(r)=TL^{(1)}_k(r)$ and impose a relation between the modes $\hat{\phi}^{(1)}_k(t)$ such that 
\begin{equation}
\left(\frac{\partial \phi^{(1)}}{\partial r}\right)_{r=0} = \sum^{N_\phi^{(1)}}_{k=0}\hat{\phi}^{(1)}_k(t)\left(\frac{dTL^{(1)}_k}{dr}\right)_{r=0} =0. \label{eq30}
\end{equation} 

\noindent A similar relation is imposed for the function $\Pi^{(1)}(t,r)$.

In the subdomain $\mathcal{D}_n$, the basis for the function $\alpha^{(n)}(t,r)$ is simply the rational Chebyshev polynomial, or $\Phi_k^{(n)} =TL_k^{(n)}(r)$. Since the asymptotic behaviors of the functions $\mu^{(n)}(t,r)$ and $\phi^{(n)}(t,r)$ are identical {(assumming that $\tilde{\phi}_0=0$)}, we have 
\begin{eqnarray}
\Xi^{(n)}_k(r)=\psi^{(n)}_k(r)=\frac{1}{2}(TL^{(n)}_{k+1}(r)-TL^{(n)}_k(r)).\label{eq31}
\end{eqnarray}

\noindent To reproduce the asymptotic behavior (\ref{eq17}), the basis for the function $\Pi^{(n)}(t,r)$ is expressed by a linear combination of the rational Chebyshev polynomials given by
\begin{eqnarray}
\chi_k^{(n)}(r) &=&  -\frac{(1+2k)}{4(3+2k)}TL_{k+2}^{(n)}(r) \nonumber \\
&&+ \frac{(1+k)}{3+2k}TL_{k+1}^{(n)}(r)-\frac{1}{4}TL_{k}^{(n)}(r). \label{eq32}
\end{eqnarray}

\noindent For the sake of completeness, we summarize the presentation of the basis functions in Table 1.

\begin{table}
	\centerline{Table 1}
	\medskip
	\begin{center}
		\centering
		\begin{tabular}[c]{ l|c|c|c }
			\hline
			&$\mathcal{D}_1$ & $\mathcal{D}_2$   \hspace{0.5cm} ... \hspace{0.5cm}   $\mathcal{D}_{n-1}$ &$\mathcal{D}_{n}$ \\
			\hline
			\hline
			$\alpha$ & $\Phi_k^{(1)}(r)$ &   & $TL_k^{(n)}(r)$ \\
			$\mu$ & $\Phi_k^{(1)}(r)$ &   & $\psi_k^{(n)}(r)$\\
			      & & $TL_k^{(l)}(r)$ &  \\
			$\phi$ & $TL_k^{(1)}(r)^{(*)}$&  & $\psi_k^{(n)}(r)$ \\
			$\Pi$  & $TL_k^{(1)}(r)^{(*)}$&  & $\chi_k^{(n)}(r)$ \\
			\hline
		\end{tabular}
	\caption{Basis functions for the spectral approximations (\ref{eq20}) - (\ref{eq23}). The asterisks indicate restriction imposed to the modes $\hat{\phi}^{(1)}_k$ and $\hat{\Pi}^{(1)}_k$ to satisfy the conditions (\ref{eq12}) and  (\ref{eq13}).}
	\end{center}
\end{table}


\subsection{Transmission conditions}

The next step is to establish the transmission conditions to guarantee that all pieces of the spectral approximations (\ref{eq20}) - (\ref{eq23}) represent the same functions but defined in each subdomain. We adopt the patching method \cite{canuto_88} that demands that a function and all their $d-1$ spatial derivatives must be continuous at the contiguous subdomains' interface. However, in the present case we introduce a slight modification consisting of imposing that all functions and their first derivatives with $r$ are continuous at the interfaces. Then, we have the following transmission conditions:
\begin{eqnarray}
\alpha^{(l)}\left(t,r^{(l)}\right)=\alpha^{(l+1)}\left(t,r^{(l)}\right), \label{eq33}\\
\nonumber \\
\left(\frac{\partial \alpha}{\partial r}\right)_{r^{(l)}}^{(l)}=\left(\frac{\partial \alpha}{\partial r}\right)_{r^{(l)}}^{(l+1)}, \label{eq34}\\
\nonumber \\
\mu^{(l)}\left(t,r^{(l)}\right)=\mu^{(l+1)}\left(t,r^{(l)}\right), \label{eq35}\\
\nonumber \\
\left(\frac{\partial \mu}{\partial r}\right)_{r^{(l)}}^{(l)}=\left(\frac{\partial \mu}{\partial r}\right)_{r^{(l)}}^{(l+1)}, \label{eq36}\\
\nonumber \\ \phi^{(l)}\left(t,r^{(l)}\right)=\phi^{(l+1)}\left(t,r^{(l)}\right), \label{eq37}\\
\nonumber \\
\left(\frac{\partial \phi}{\partial r}\right)^{(l)}_{r^{(l)}}=\left(\frac{\partial \phi}{\partial r}\right)_{r^{(l)}}^{(l+1)}, \label{eq38}
\end{eqnarray}

\noindent with $l=1,2,..,n-1$, and the same conditions for the function $\Pi(t,r)$. Two remarks are in order. First, the number of transmission conditions restricts the number of collocation or grid points in the subdomains; and second, it is possible to generate a stable evolution code suppressing the condition (\ref{eq36}) if one follows the imposition of the patching method rigorously.

\subsection{The grid or collocation points}

The collocation points located in the computational subdomains $-1 \leq\xi_j^{(l)} \leq 1$ are mapped into the physical domain according to {the inverse of the argument in the rational  Chebyshev  polynomials given by (\ref{eq26}) }
\begin{eqnarray}
r^{(l)}_j=\frac{b^{(l)}-L_0\xi_j^{(l)}}{\xi_j^{(l)}-a^{(l)}}, \label{eq39}
\end{eqnarray}

\noindent with $a^{(l)},b^{(l)}$ given by the relations (\ref{eq27}) and (\ref{eq28}), respectively. By designating $N^{(l)}$ the truncation order of any field function ($N_\alpha^{(l)}, N_\mu^{(l)}, N_\phi^{(l)}$ or $N_\Pi^{(l)}$), the total number collocation points, $\xi^{(l)}_j$ in the computational subdomains must be  
\[
\underbrace{\sum_{l=1}^n N_\mu^{(l)}+n}_{\substack{\text{number of}\\\text{modes}}} - \underbrace{2(n-1)}_{\substack{\text{transmission}\\\text{conditions}}} = 
\underbrace{\sum_{l=1}^n N_\mu^{(l)}-n+2}_{\substack{\text{number of collocation}\\\text{points}}},
\]
\noindent where $n \geq 2$. We have distributed the collocation points as shown in Table 2.

\begin{table}
	\centerline{Table 2}
	\medskip
	\begin{center}
		\centering
		\begin{tabular}[c]{ l|c|c|c|c|r }
			\hline
			&$\mathcal{D}_1$ & $\mathcal{D}_2$ &  \hspace{0.5cm} ... \hspace{0.5cm}  & $\mathcal{D}_{n-1}$ &$\mathcal{D}_{n}$ \\
			\hline
			\hline
			$\alpha$ & $N^{(1)}_\alpha$ & $N^{(2)}_\alpha-1$ & ... & $N^{(n-1)}_\alpha-1$ & $N^{(n)}_\alpha$ \\
			$\mu$ & $N^{(1)}_\mu$ & $N^{(2)}_\mu-1$ & ... & $N^{(n-1)}_\mu-1$ & $N^{(n)}_\mu$\\
			$\phi$ & $N_\phi^{(1)}$ & $N_\phi^{(2)}-1$ & ... & $N_\phi^{(n-1)}-1$ & $N_\phi^{(n)}$ \\
			$\Pi$ & $N_\Pi^{(1)}$ & $N_\Pi^{(2)}-1$ & ... & $N_\Pi^{(n-1)}-1$ & $N_\Pi^{(n)}$ \\
			\hline
		\end{tabular}
		\caption{Distribution of the collocation points in each subdomains according to corresponding truncation orders.}
	\end{center}
\end{table}

We have chosen two sets of Chebyshev-Gauss-Lobatto points; one for the metric functions $\alpha(t,r), \mu(t,r)$, and another for the matter functions $\phi(t,r), \Pi(t,r)$. These sets can have a distinct number of points in each subdomain,  where we have excluded the origin from the first set since the spectral approximations for $\alpha(t,r)$ and $\mu(t,r)$ fix these functions at $r=0$. The following  collocation points characterize the first set:
\begin{eqnarray}
\xi_j^{(1)} &=& \cos\left(\frac{j \pi}{N_\mu^{(1)}+1}\right),\;\;j=1,2,..,N_\mu^{(1)}, \label{eq40}\\
\nonumber \\
\xi_j^{(l)} &=& \cos\left(\frac{j \pi}{N_\mu^{(l)}-1}\right),\;\;j=1,2,..,N_\mu^{(l)}-1,\label{eq41}
\end{eqnarray}

\noindent for $l=2,..,n-1$, and for the last subdomain
\begin{eqnarray}
\xi_j^{(n)} &=& \cos\left(\frac{j \pi}{N_\mu^{(n)}}\right),\;\;j=1,2,..,N_\mu^{(n)}, \label{eq42}
\end{eqnarray}

\noindent where we assume that $N_\alpha^{(l)}=N_\mu^{(l)}$. The second set of collocation points is given by 
\begin{eqnarray}
\xi_j^{(l)} = \cos\left(\frac{j \pi}{N_\phi^{(l)}}\right),\;\; j=1,2,..,N_\phi^{(l)}, \label{eq43}
\end{eqnarray}

\noindent for the first and the last subdomains, $l=1,n$, respectively; and the for the interior subdomains, $l=2,..,n-1$, we have

\begin{eqnarray}
\xi_j^{(l)} &=& \cos\left(\frac{j \pi}{N_\phi^{(l)}-1}\right),\;\; j=1,2..,N_\phi^{(l)}-1. \label{eq44}
\end{eqnarray} 

\noindent Here, $N^{(l)}_\Pi=N^{(l)}_\phi$. We would like to make some pertinent and brief comments. The sets of collocation points (\ref{eq40}) - (\ref{eq42}) and (\ref{eq43}) - (\ref{eq44}) have identical structure; the exception is that the origin is included in the second set.  We have placed the interfaces $r^{(1)},r^{(2)},..,r^{(n-1)}$ as belonging to the corresponding subsequent subdomains, $\mathcal{D}_2,\mathcal{D}_3,..,\mathcal{D}_n$, respectively, as implemented in Ref. \cite{alcoforado}. 

\subsection{Implementing the GC domain decomposition method}


The goal of any spectral method is to approximate a set of evolution partial differential equations by a finite set of ordinary differential equations. By considering the system formed by the equations  (\ref{eq2}), (\ref{eq3}), (\ref{eq6}) and (\ref{eq7}), we approximate them by sets of dynamical systems for the values $\phi_j^{(l)}(t)$, $\Pi_j^{(l)}(t)$ and $\mu_j^{(l)}(t)$ with $l=1,2,..,n$, or equivalently for the corresponding modes. The field equation $(\ref{eq2})$ results in a set of algebraic equations for the modes $\hat{\alpha}_k^{(l)}(t)$. 

We describe in the sequence the evolution scheme of the multidomain GC method (see also Refs. \cite{alcoforado,barreto_19,crespo_19}). By substituting the spectral approximations (\ref{eq20}) - (\ref{eq23}) into the field equations (\ref{eq2}), (\ref{eq3}), (\ref{eq6}) and (\ref{eq7}) we obtain the corresponding residual equations. For instance, considering  Eq. (\ref{eq2}), the residual equations in each subdomain become
\begin{eqnarray}
\mathrm{Res}_\alpha^{(l)}(t,r) =\left(\alpha_{,r}-\mu_{,r}+\displaystyle{\frac{1}{r}}(1-\mathrm{e}^{2\mu})+r\mathrm{e}^{2\mu}V(\phi)\right)^{(l)},\label{eq45} \nonumber \\
\end{eqnarray}

\noindent with $l=1,2,..,n$. These equations do not vanish due to the approximations established for the metric and scalar field functions. We have followed the prescription of the Collocation method for updating the modes $\alpha^{(l)}_k(t)$.  It means requiring the above residual equations to vanish at the collocation points located in each subdomain. Then   
\begin{eqnarray}
(\alpha_{,r})_j^{(l)}=(\mu_{,r})^{(l)}_j-\frac{1}{r^{(l)}_j}\left(1-\mathrm{e}^{2\mu_j^{(l)}}\right)-r_j^{(l)}\mathrm{e}^{2\mu_j^{(l)}}V(\phi_j^{(l)}), \nonumber  \label{eq46} \\
\end{eqnarray}

\noindent where $r_j^{(l)}$ are the collocation points in the physical domain given by Eq. (\ref{eq39}). With the values $(\mu_{,r})^{(l)}_j, \mu_j^{(l)}, \phi_j^{(l)}$ we can determine the values $(\alpha_{,r})_j^{(l)}$, and, as we are going to show, the modes $\hat{\alpha}_k^{(l)}$.

For convenience, it will be helpful to introduce a new notation for the values and modes of all metric and scalar functions defined in each subdomain. Schematically, we have

\[\mu_j^{(l)}\,\rightarrow\, \mu_j,\;\;\hat{\mu}_j^{(l)}\,\rightarrow\, \hat{\mu}_j  \] 
\[\phi_j^{(l)}\,\rightarrow\, \phi_j,\;\;\hat{\phi}_j^{(l)}\,\rightarrow\, \hat{\phi}_j,\,\,\mathrm{and\,so\,on.}  \]

\noindent In this case, we determine the values $(\mu_{,r})_j$ in the following matrix form 
\begin{eqnarray}
\begin{pmatrix}
(\mu_{,r})_1\\
(\mu_{,r})_2\\
\vdots\\
(\mu_{,r})_{N_\mu^{(1)}}\\
(\mu_{,r})_{N_\mu^{(1)}+1}\\
\vdots \\
(\mu_{,r})_{N_T}	
\end{pmatrix}
=\mathbb{DM}\,
\begin{pmatrix}
\hat{\mu}_1\\
\hat{\mu}_2\\
\vdots\\
\hat{\mu}_{N^{(1)}_\mu+1}\\
\hat{\mu}_{N^{(1)}_\mu+2}\\
\vdots \\
\hat{\mu}_{\hat{N}_T}	
\end{pmatrix}, \label{eq47}
\end{eqnarray}

\noindent where $N_T=\sum^n_{l=1}\,N_\mu^{(l)}-n+2$ and $\hat{N}_T=\sum^n_{l=1}\,N_\mu^{(l)}+n$ are the total number of values and modes, respectively and

\[\mathbb{DM}_{jk} = \left(\frac{\partial \Xi_k}{\partial r}\right)_{r_j}.\]

\noindent Notice that the first $N_\mu^{(1)}$ entries of matrix on the LHS are the values $(\mu_{,r})_j$ in the first subdomain; the subsequent entries correspond to the values $(\mu_{,r})_j$ belonging to the other subdomains. Therefore, all values and modes are expressed in column vectors, and we have calculated all derivatives of the metric and scalar field functions similarly.

In order to determine the modes $\hat{\alpha}_k$ from the values $(\alpha_{,r})_j$ obtained from Eq. (\ref{eq46}), we have to include the transmission conditions (\ref{eq33}) and (\ref{eq34}). Since we have more modes than values, we can establish the matrix equation
\begin{eqnarray}
\begin{pmatrix}
(\alpha_{,r})_1\\
(\alpha_{,r})_2\\
\vdots\\
(\alpha_{,r})_{N_T}\\
0\\
\vdots\\
0	
\end{pmatrix}
=\mathbb{DAL}\,
\begin{pmatrix}
\hat{\alpha}_1\\
\hat{\alpha}_2\\
\vdots\\
\hat{\alpha}_{N_T}\\
\hat{\alpha}_{N_T+1}\\
\vdots \\
\hat{\alpha}_{\hat{N}_T}	
\end{pmatrix}, \label{eq48}
\end{eqnarray}

\noindent where 

\[\mathbb{DAL}_{jk} = \left(\frac{\partial \Phi_k}{\partial r}\right)_{r_j},\]


\noindent for all $j=1,2,..,N_T, \, k=1,2,..,\hat{N}_T$. The components $\mathbb{DAL}_{jk}$ with $j = N_T+1,..,\hat{N}_T$ are fixed by the transmission conditions $(\ref{eq33})$ and $(\ref{eq34})$ expressed in terms of the modes $\hat{\alpha}_k$. Further, the zeros entries on the LHS correspond to the $2(n-1)$ of these conditions. Then, after {solving the linear system} using standard methods, we can update modes $\hat{\alpha}_k$ enforcing the fulfillment of the transmission conditions.

We have followed the same procedure to update the modes $\hat{\mu}_k(t)$. After vanishing the residual equation at the collocation points in each domain and introducing the new notation, we have
\begin{eqnarray}
\displaystyle{(\mu_{,t})_j = \frac{1}{2}r_j\mathrm{e}^{\alpha_j-\mu_j}\Pi_j (\phi_{,r})_j}. \label{eq49}
\end{eqnarray}

\noindent With these values, we can calculate the values $\mu_j$ at the next time level. Now, we obtain the modes $\hat{\mu}_k$ using the matrix equation 
\begin{eqnarray}
\begin{pmatrix}
\mu_1\\
\mu_2\\
\vdots\\
\mu_{N_T}\\
0\\
\vdots\\
0	
\end{pmatrix}
=\mathbb{M}\,
\begin{pmatrix}
\hat{\mu}_1\\
\hat{\mu}_2\\
\vdots\\
\hat{\mu}_{N_T}\\
\hat{\mu}_{N_T+1}\\
\vdots \\
\hat{\mu}_{\hat{N}_T}	
\end{pmatrix}, \label{eq50}
\end{eqnarray}

\noindent where the $2(n-1)$ zeros entries in the LHS correspond to the transmission conditions (\ref{eq35}) and (\ref{eq36}) expressed in terms of the modes $\hat{\mu}_k$. The components of the square matrix $\mathbb{M}$ are

\[\mathbb{M}_{jk} = \Xi_k(r_j), \]

\noindent $j=1,2,..,\hat{N}_T,\,k=1,2,..,\hat{N}_T$, and those components $\mathbb{M}_{jk}$ with $j=N_T+1,..,\hat{N}_T$ are fixed by the transmission conditions. Thus, 
{again we solve the linear system} to update the modes $\hat{\mu}_k$. {We note here that the linear system is solved at the initial time, reusing the saved information at any time. Thus the computational performance is substantially improved, 
only requiring storage in memory.}

Finally, we have envisaged a similar procedure to update the values and modes associated with the scalar field $\phi(t,t)$ and $\Pi(t,r)$. The main difference refers to the use of a new set of collocation points (cf. Eqs. (\ref{eq43}) and (\ref{eq44})) that includes the origin. For the sake of completeness, after vanishing the residual equations related to (\ref{eq6}) and (\ref{eq7}), we have
\begin{eqnarray}
(\phi_{,t})_j &=& \mathrm{e}^{\alpha_j-\mu_j}\,\Pi_j, \label{eq51}\\
\nonumber \\
(\Pi_{,t})_j &=& \big[\left(\mathrm{e}^{\alpha-\mu}\phi_{,r}\right)_{,r}\big]_j + \frac{2}{r_j}\mathrm{e}^{\alpha_j-\mu_j}(\phi_{,r})_j \nonumber \\
&& - \mathrm{e}^{\alpha_j-\mu_j}\left(\frac{dV}{d\phi}\right)_j. \label{eq52}
\end{eqnarray}


\noindent We have calculated all values present on the RHS of both equations as before. In particular, for the term $\phi_{,r}/r$ at $r=0$, it is necessary expressing it explicitly in terms of the modes $\hat{\alpha}_k$. Due to the choice of the basis function $\Phi_k(r)$ given by Eq. (\ref{eq29}), we guarantee the regularity at the origin. Eqs. (\ref{eq51}) and (\ref{eq52}) allow for updating the values $\phi_j$, $\Pi_j$, therefore providing the modes $\hat{\phi}_k$ and $\hat{\Pi}_k$ in the next time level, respectively, where the transmission conditions are inserting in the same fashion as in Eq. (\ref{eq50}). 

In closing this Section, we have presented the evolution code's primary machinery. Besides, we point out that there is no unique way of establishing the transmission conditions for the domain decomposition method applied to hyperbolic problems \cite{kopriva_86,kopriva_89}. The delicate problem is to choose the proper way the information is transmitted through several subdomains, which is crucial to provide a stable code. In establishing the transmission conditions (\ref{eq35}) - (\ref{eq36}) for the dynamical functions, we have adopted their simplest form and enforcing that each interface constitutes a collocation point of the subsequent subdomain. As we will show in the next Section, the code is stable and converge exponentially no matter the chosen error measure.

\section{Code validation: numerical results}


We present the numerical tests to validate the domain decomposition  GC algorithm by investigating the influence of increasing the resolution in each subdomain and the number of subdomains. In {almost} all numerical experiments, we integrate the resulting dynamical system with a fourth-order Runge-Kutta integrator with time step sizes ranging from $10^{-4}$ to {$10^{-5}$. We have to reduce the time step size when increasing the spatial resolution, and roughly the time step is proportional to $1/N^2$, where $N$ is the total truncation order. The map parameter has some influence on the stepsize, as well. We have not found any other problem than the time step limit for each spatial resolution}. Furthermore, we have adopted a strategy of dividing the intermediate computational domain covered by $-1 \leq x \leq 1$ (cf. Fig. 1) in equal parts. It means that in the case of two subdomains, for instance, $x^{(1)}=0$ implying in $r^{(1)}=L_0$; for three subdomains, we have $x^{(1)}=-1/3$ and $x^{(2)}=1/3$ yielding $r^{(1)}=L_0/2$ and $r^{(2)}=2 L_0$, respectively, and so on. The remaining information about the location of the interfaces is determined by the map parameter $L_0$. Unless stated otherwise, we adopt the following, say ``economic" rule for choosing the truncation orders that appear in the spectral approximation (\ref{eq20}) - (\ref{eq23}). Regardless the number of subdomains, we fix the following relation between the truncation orders present in Eqs. (20) - (23): $N_\Pi^{(l)}=N_\phi^{(l)}$ and $N_\mu^{(l)}=N_\alpha^{(l)}$. For two subdomains, we first fix $N^{(1)}_\phi$ then $N^{(2)}_\phi=1.3 N^{(1)}_\phi$, and $N^{(1)}_\alpha=1.2N^{(1)}_\phi,\;N^{(2)}_\alpha=1.4N^{(1)}_\phi$. For three subdomains, we have, after fixing $N^{(1)}_\phi$, the truncation orders $N^{(2)}_\phi=N^{(1)}_\phi+1$, and $N^{(3)}_\phi=1.2 N^{(1)}_\phi$; and $N^{(1)}_\alpha=1.1N^{(1)}_\phi,\;N^{(2)}_\alpha=1.1N^{(1)}_\phi+1,\;N^{(3)}_\alpha=1.3N^{(1)}_\phi+1$.


%
\begin{figure}[htb]
\vspace{-1cm}	
	\includegraphics[width=9.cm,height=8.cm]{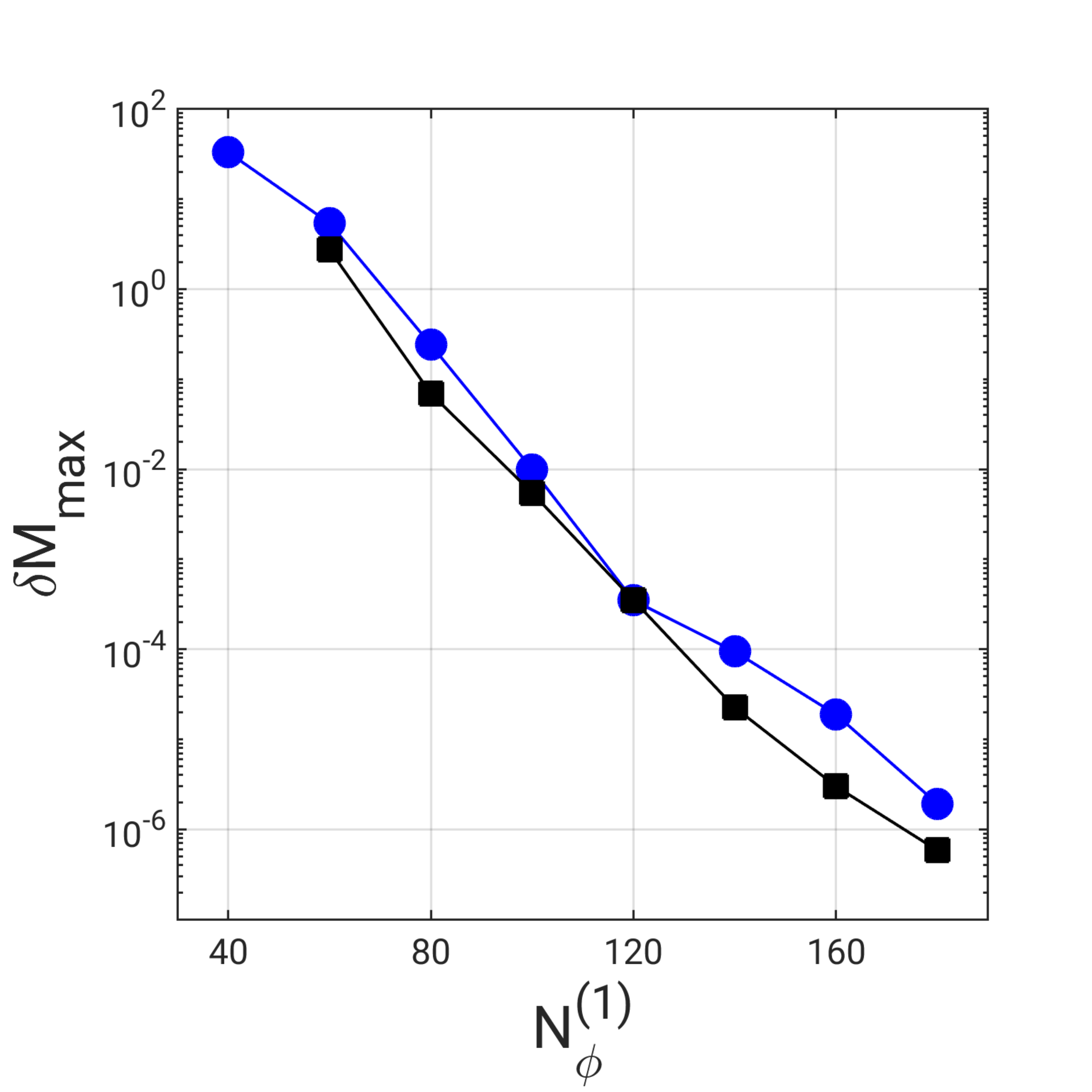}
	\vspace{-0.3cm}
	\caption{Exponential decay of the maximum relative deviation $\delta M$ for $L_0=5$ (blue circles) and $L_0=10$ (black boxes) for a code with two subdomains. $N_\phi^{(1)}$ is the truncation order in the first domain taken as a reference. For $N_\phi^{(1)}=180$, the maximum error is about $10^{-6}\%$. The final time of integration is $t=10$.}
\end{figure}

\begin{figure}[htb]
	\vspace{-0.5cm}
	\includegraphics[width=9.cm,height=8.cm]{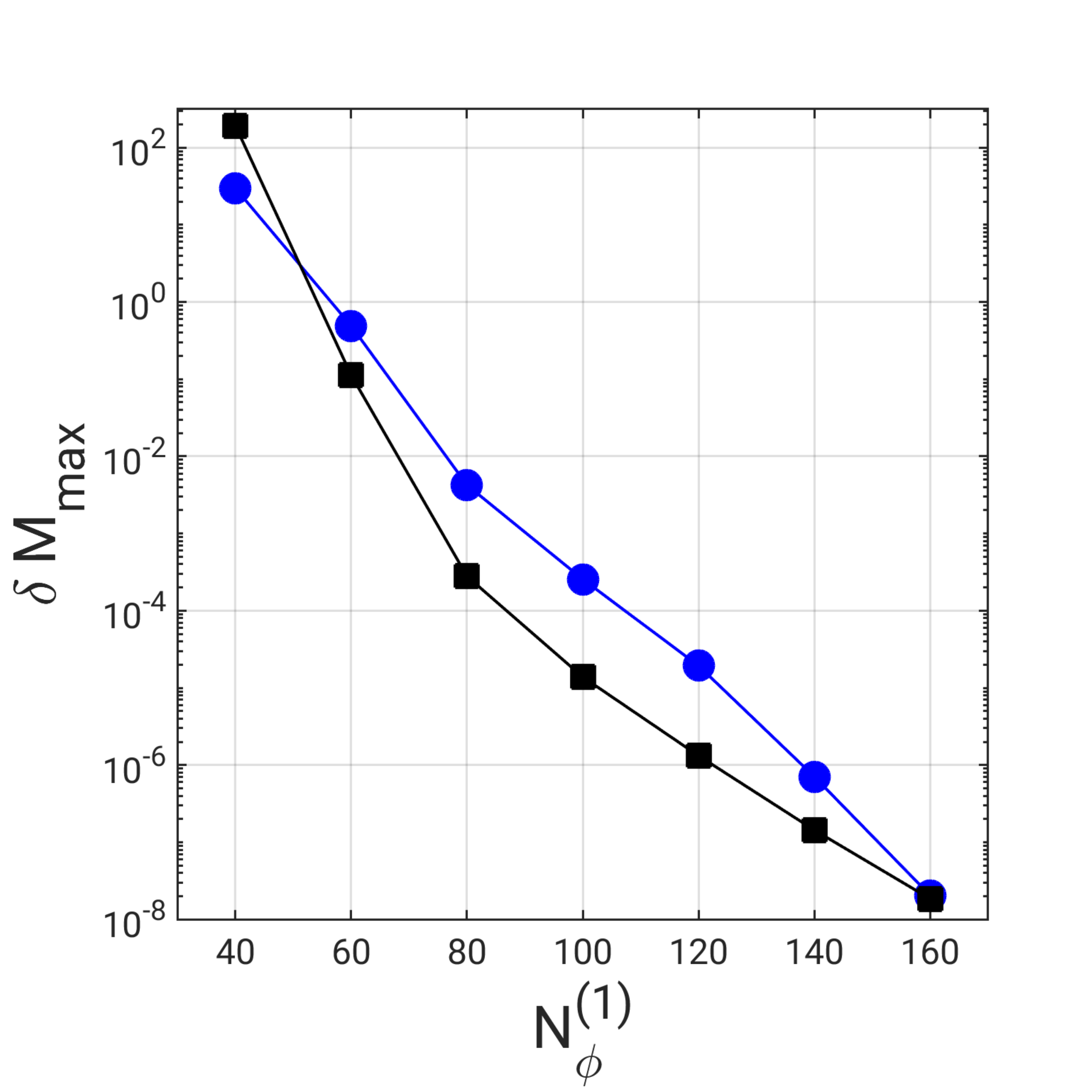}
	\vspace{-0.3cm}
	\caption{Exponential decay of the maximum relative deviation $\delta M_{\mathrm{max}}$ for $L_0=5$ (blue circles), and $L_0=10$ (black boxes) produced by a code with three subdomains. $N_\phi^{(1)}$ is the truncation order in the first domain taken as a reference. The final time of integration is $t=10$.}
\end{figure}

The numerical experiments to validate the present algorithm start with the initial data taken from the Alcubierre's book \cite{alcubierre}

\begin{equation}
\phi_0(r) = A_0 r^2\mathrm{e}^{-(r-r_0)^2/\sigma^2}, \label{eq53}
\end{equation}

\noindent where $\sigma=1.0$, $r_0=5.0$ and $A_0$ is the initial amplitude left as a free parameter. We also assume that $\Pi(0,r)=\Pi_0(r)=0$, and the initial function $\mu(0,r)=\mu_0(r)$ is determined after solving the Hamiltonian constraint (\ref{eq4}). In all numerical tests, we have integrated the system until $t=10$.

The first numerical test consists of verifying the conservation of the ADM mass. We calculate the relative deviation, $\delta M$, according to 
\begin{eqnarray}
\delta M (t)= \frac{|M_{ADM}(0)-M_{ADM}(t)|}{M_{ADM}(0)} \times 100,\label{eq54}
\end{eqnarray}

\noindent where $M_{ADM}(0)$ is the initial ADM mass and  $M_{ADM}(t)$ is its corresponding numerical value at an instant $t$. In all simulations, we choose the initial scalar field amplitude $A_0=0.008$ that is close to the formation of an apparent horizon ($A_0  \gtrsim 0.0082$), therefore providing a typical nonlinear evolution. 

\begin{figure}[htb]
    \includegraphics[width=8.0cm,height=7.5cm]{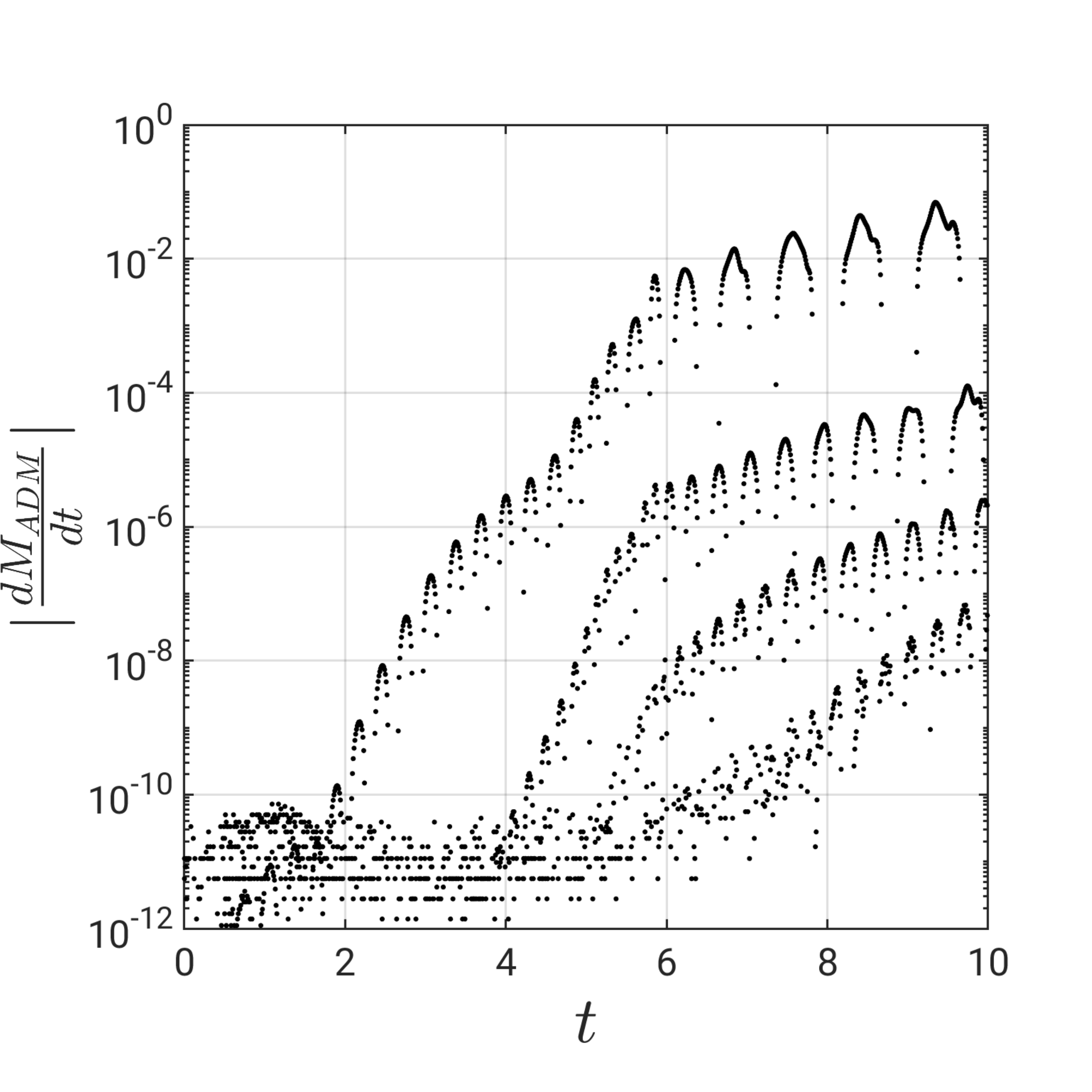}
	\vspace{-0.1cm}
	\includegraphics[width=8.cm,height=7.5cm]{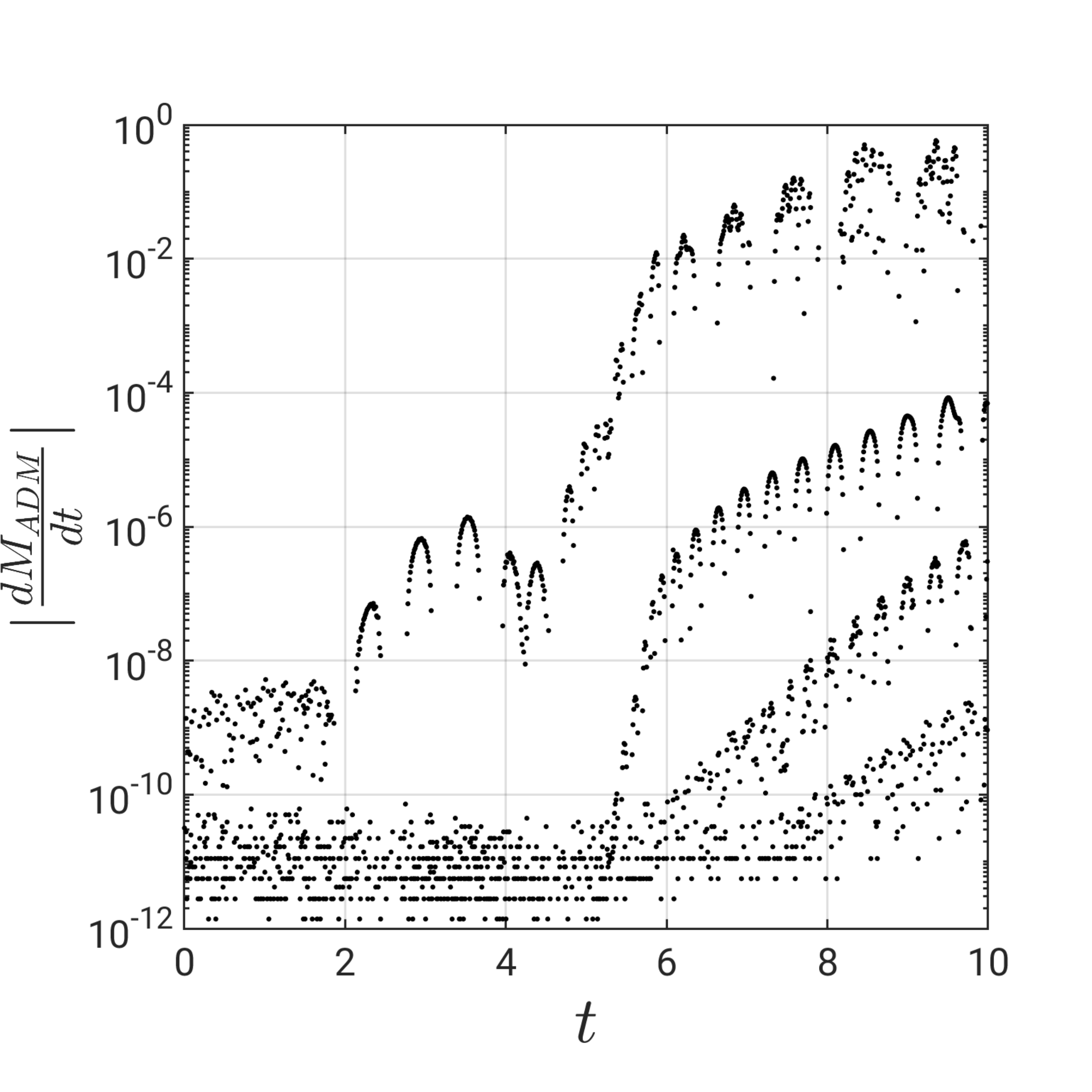} 
	\caption{Illustration of the conservation of the ADM mass produced by the code with two (upper panel) and three sudbomains (lower panel). In the log-linear plots of $|d M_{ADM}/d t|$, we have set $L_0=5$. {The selected resolutions are $N^{(1)}_\phi=60,100,140,180$ (two subdomains) and $N^{(1)}_\phi=40,80,120,180$ (three subdomains).}}
\end{figure}







We calculate the ADM mass using the integral formula (\ref{eq4}) taking advantage of the scheme shown in Fig. 1 by splitting the integral to a sum of integrals evaluated in the corresponding computational subdomains covered by $-1 \leq \xi^{(l)} \leq 1$, $l=1,2..,n$, or
\begin{eqnarray}
M_{ADM} = \sum_{l=1}^n\,\int_{-1}^{1}\left[\frac{r^2}{4} \mathrm{e}^{-2 \mu}\left(\Pi^2+\phi_{,r}^2\right)\right]^{(l)}\left(\frac{dr}{d \xi}\right)^{(l)} d\xi^{(l)}, \nonumber \\ \label{eq55}
\end{eqnarray}

\noindent where $n$ is the number of subdomains, the factor $dr/d \xi$  is calculated using the relation (\ref{eq39}), and we have considered a massless scalar field. We further approximate each integral with quadrature formulae \cite{fornberg} as 
\begin{eqnarray}
\int_{-1}^{1}\left[\frac{r^2}{4} \mathrm{e}^{-2 \mu}\left(\Pi^2+\phi_{,r}^2\right)\right]^{(l)}&&\left(\frac{dr}{d \xi}\right)^{(l)} d\xi^{(l)} \nonumber \\
&&\approx \sum_{k=0}^{N_q}\,(...)^{(l)}_k w_k^{(l)}. \label{eq56}
\end{eqnarray}

\noindent In the above expression, $w_k^{(l)}$ represents the weights, given by 

\[w_k^{(l)} = \frac{1}{N_q^{(l)} \left(1+N_q^{(l)}\right)}\frac{2}{P_{N_q^{(l)}}\left(\xi_k^{(l)}\right)},\]

\noindent where $N_q^{(l)}=1.5 N_\phi^{(l)}$ is the quadrature truncation order defined in each subdomain, and $(...)_k$ denote the values of the integrand at the quadrature collocation points assumed to be the Legendre-Gauss-Lobatto points \cite{fornberg}. Also, in evaluating the integrals, we have used the spectral approximations (\ref{eq20}) - (\ref{eq23}).


{We present in Fig. 2 the convergence of the maximum value of $\delta M$ with a two subdomain code by increasing the resolution represented by the first subdomain's truncation order, $N_\phi^{(1)}$ as a reference}. We have tested two values of the map parameter, $L_0=5$ and $L_0=10$ (circles and boxes, respectively), fixing the interfaces between the subdomains at $r^{(1)}=5$ and $10$ respectively. The maximum deviation, $\delta M_{\mathrm{max}}$, decays exponentially at similar rates, but for $L_0=10$, the exponential decay appears to be slightly better. For the resolution with $N^{(1)}_\phi=180$, the maximum relative deviation in the ADM mass is about $10^{-6}\%$. Although not exhibited here, smaller map parameter values produce a slower exponential decay of $\delta M_{\mathrm{max}}$. We list some of the total integration times: for $N_\phi^{(1)}=60$, the stepsize was $10^{-4}$, and the simulation lasted approximately $94 s$; for $N_\phi^{(1)}=100$ with stepsize of $2 \times 10^{-5}$ the total integration time was about $17$ minutes. The longest integration time was approximately $3.3$ hours for $N_\phi^{(1)}=180$ and stepsize $5 \times 10^{-6}$. We have used a computer with a Core $i7-8700K$ processor and $64$ Gb of RAM.

\begin{figure}[htb]
	\includegraphics[width=8.5cm,height=7.5cm]{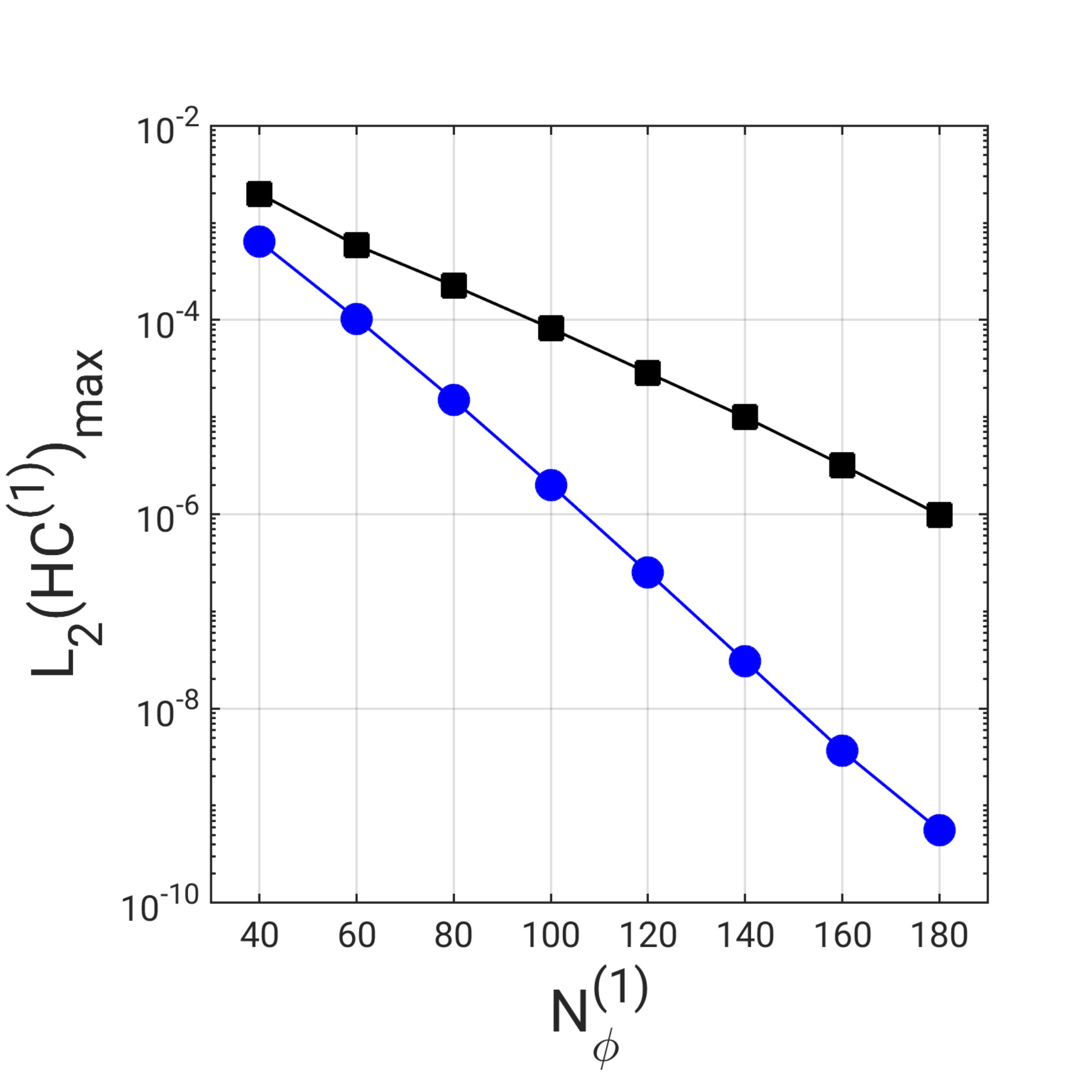}
	\\
	\includegraphics[width=8.5cm,height=7.5cm]{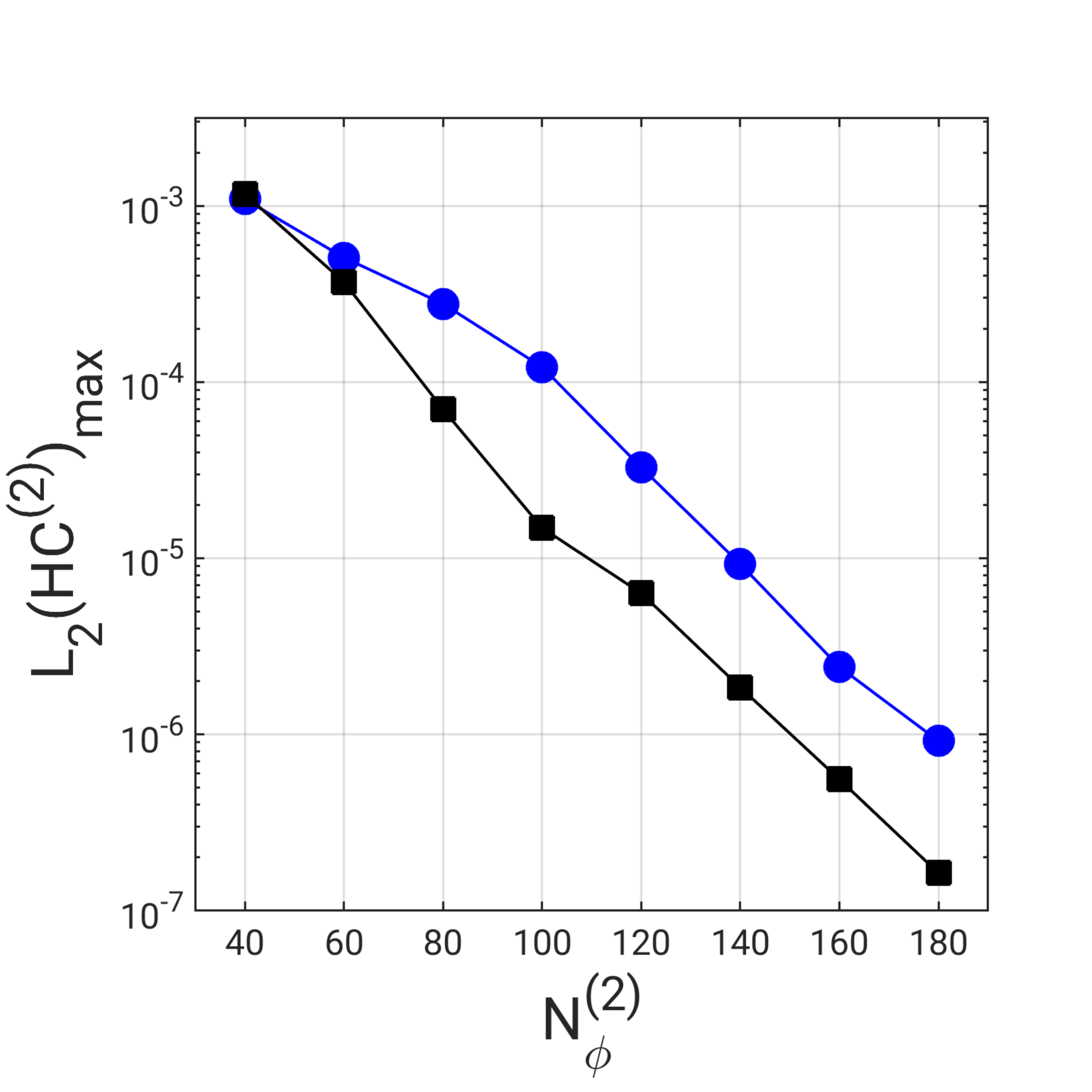}
	\caption{Decay {of the maximum value $L_2$-error at any time in the evolution,} associated with the Hamiltonian constraint in the first (upper panel) and second (lower plot) subdomains. We have used two values of the map parameter: $L_0=5$ (blue circles),  and $L_0=10$ (black boxes). $N_\phi^{(1)}$ and $N_\phi^{(2)}{=1.3N_\phi^{(1)}}$ are the truncation orders of the scalar field in the first and second subdomains, respectively.}
\end{figure}

We repeat the same numerical test in Fig. 3 with three subdomains taking  the same values of map parameter, $L_0=5$ and $L_0=10$, which places the subdomains interfaces at $r^{(1)}=2.5$, $r^{(2)}=10$, and $r^{(1)}=5$, $r^{(2)}=20$, respectively. According to the plots in Fig. 3, the exponential decay is evident for $L_0=5$ inside the range $40 \leq N^{(1)}_\phi \leq 160$, whereas for $L_0=10$ the exponential decay takes place in $80 \leq N^{(1)}_\phi \leq 160$. As a consequence, the maximum error reaches approximately the same value of $10^{-8} \%$ when  $N^{(1)}_\phi = 160$ for both values of $L_0$.  It means that the ADM mass is conserved up to one part in $10^{10}$. In this case, some of the integration times are: for $N_\phi^{(1)}=60$ with stepsize of $10^{-4}$, the simulation lasted approximately $2.7$ minutes; for $N_\phi^{(1)}=100$ and stepsize $2 \times 10^{-5}$, the total integration time was about $25$ minutes. The longest integration time was approximately $4.6$ hours for $N_\phi^{(1)}=160$ and stepsize $5 \times 10^{-6}$. {We have not noticed a substantial difference in the integration time with a two-domain code with 120 modes in each domain and a three-domain code with 80 spectral modes in each domain. For an integration to $t=10$ and a step size of $2\times 10^{-5}$ we needed approximately $50$ minutes. However, for higher resolution and increasing the number of subdomains, we obtain more sparse matrices. Then, we expect that with more subdomains, the integration tends to be faster.} 

Another way of illustrating the effect of increasing the truncation order in the conservation of the ADM mass is to plot the numerical values of the derivative of the numerical ADM mass with respect to time as

\[\frac{d M_{ADM}}{dt} \approx \frac{1}{2h} \left(M_{ADM}(t+h)-M_{ADM}(t-h)\right),\]

\noindent  where $h$ is the stepsize. {In Fig. 4} we present the results using two (upper panel) and three subdomains (lower panel) codes with map parameter $L_0=5$.  The selected resolutions are $N^{(1)}_\phi=60, 100, 140, 180$ and $N^{(1)}_\phi=40, 80, 120, 160$, for two and three subdomains codes, respectively. 

{Although we have exhibited an exponential convergence associated with the calculation of the ADM mass, it is essential to comment on the growth of the error displayed in Fig. 4. The numerical determination of the ADM mass is very delicate since it is a quantity that lives in the spatial infinity. Covering all spatial domains is one of the virtues of the present code, which allows a precise evaluation of the ADM mass. However, the collocation points are more sparse to a long distance from the origin for small resolution. It means that when the scalar field pulse reaches a region with low resolution, errors are introduced and strongly influence the value of the ADM mass. For higher resolutions, the pulse can reach long distances but with a decreasing amplitude producing, as a consequence, a slower increase in the error in the ADM mass as indicated by Fig. 4. In any case, the ADM mass error growth does not imply any instability, as we have verified. We refer to Ref. \cite{wtichy} as an illustration of the sensitiveness of the ADM mass calculation in another system, but using spectral methods.}.

The next numerical test is the validity and convergence of the Hamiltonian constraint (\ref{eq4}). We proceed by calculating the $L_2$-error associated with the Hamiltonian constraint at each subdomain. Then, we may write
\begin{eqnarray}
L_2(HC^{(l)}) = \sqrt{\frac{1}{2}\int_{-1}^1\,(\mathcal{H}^{(l)})^2 d\xi^{(l)}}, \label{eq57}
\end{eqnarray}

\noindent with $l=1,2,..,n$, and the integral is calculated using a quadrature formula.
\begin{figure}[htb]
\vspace{-1cm}
		\includegraphics[width=7.cm,height=6.cm]{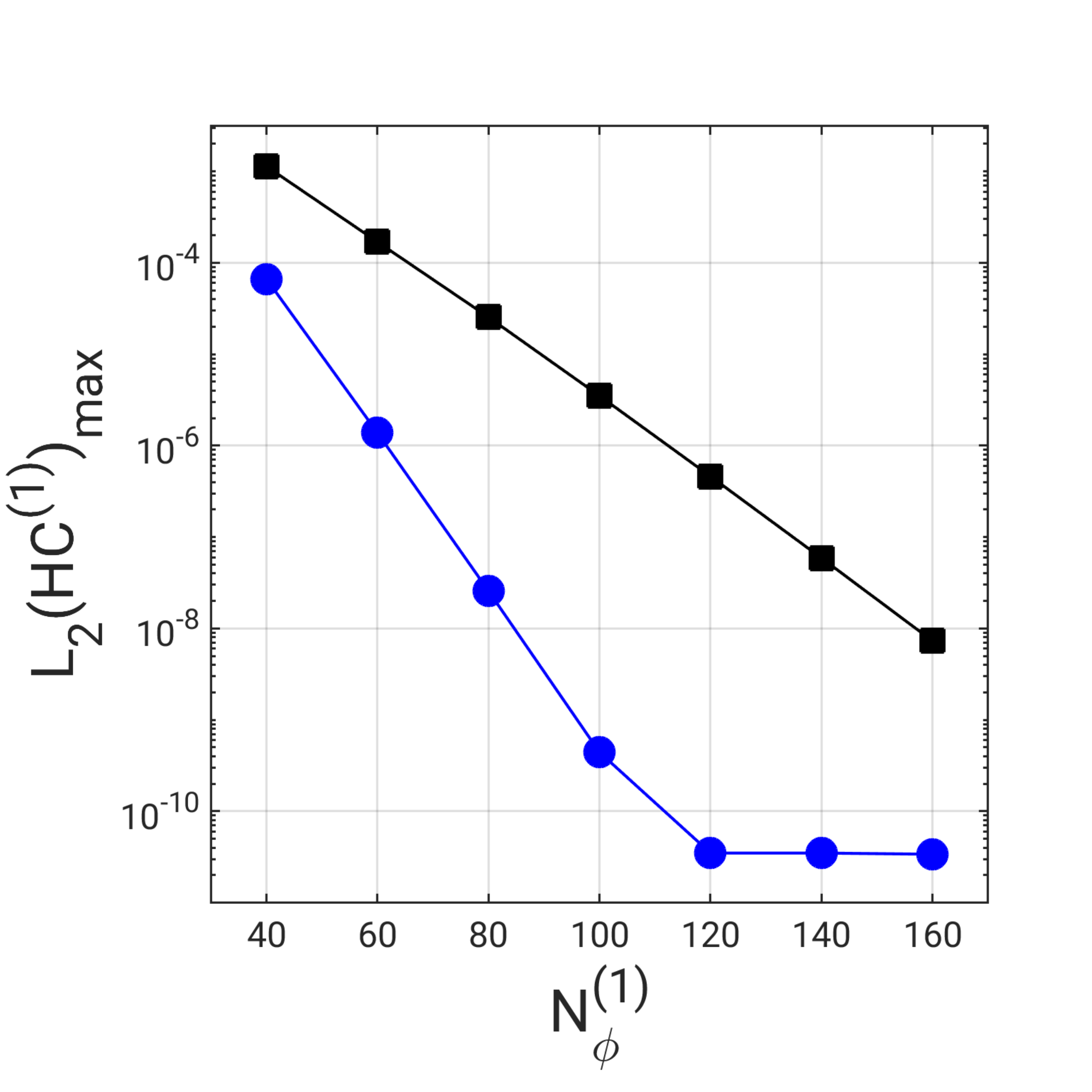}
	\vspace{-0.1cm}
	\includegraphics[width=7.cm,height=6.cm]{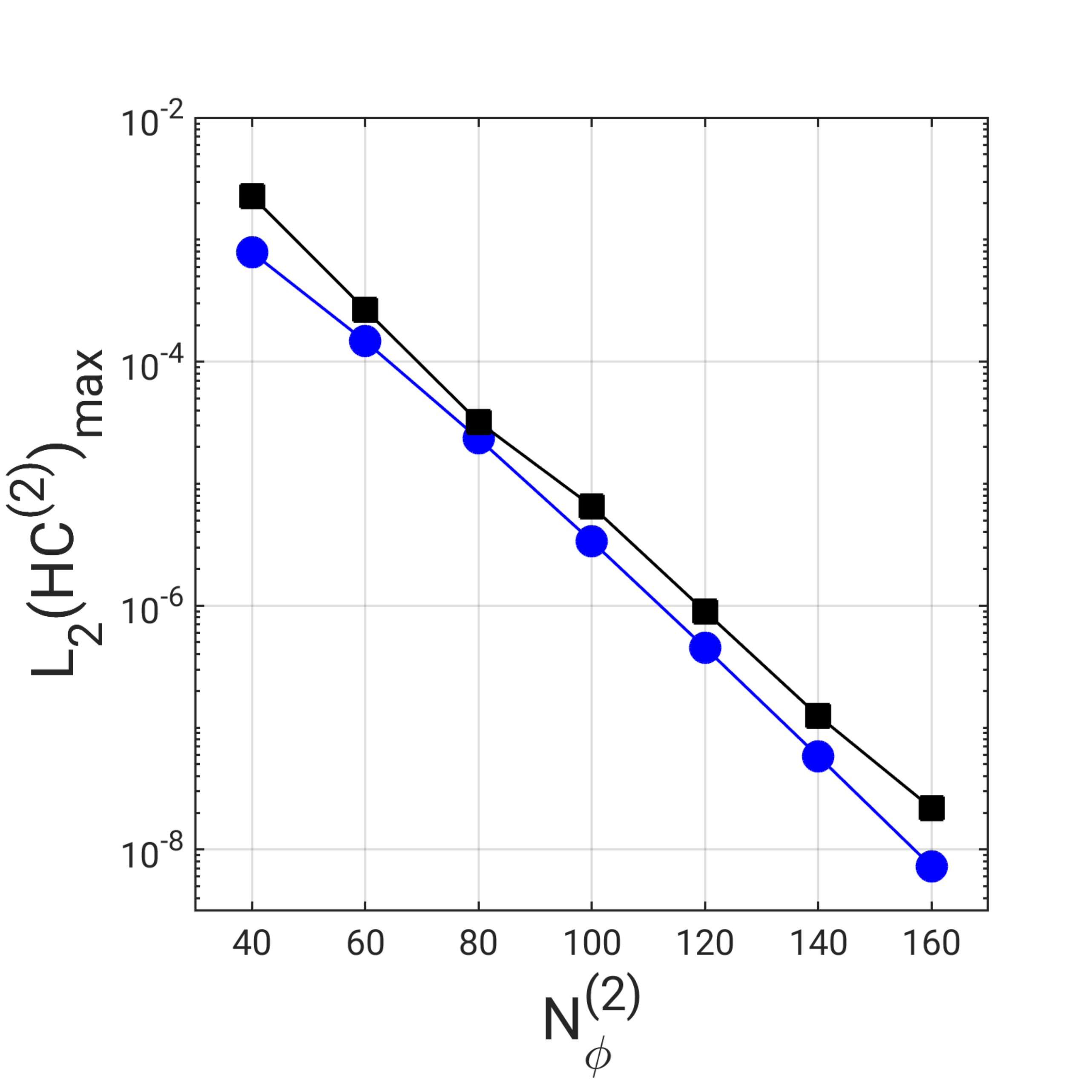}
	\vspace{-0.1cm}
	\includegraphics[width=7.cm,height=6.cm]{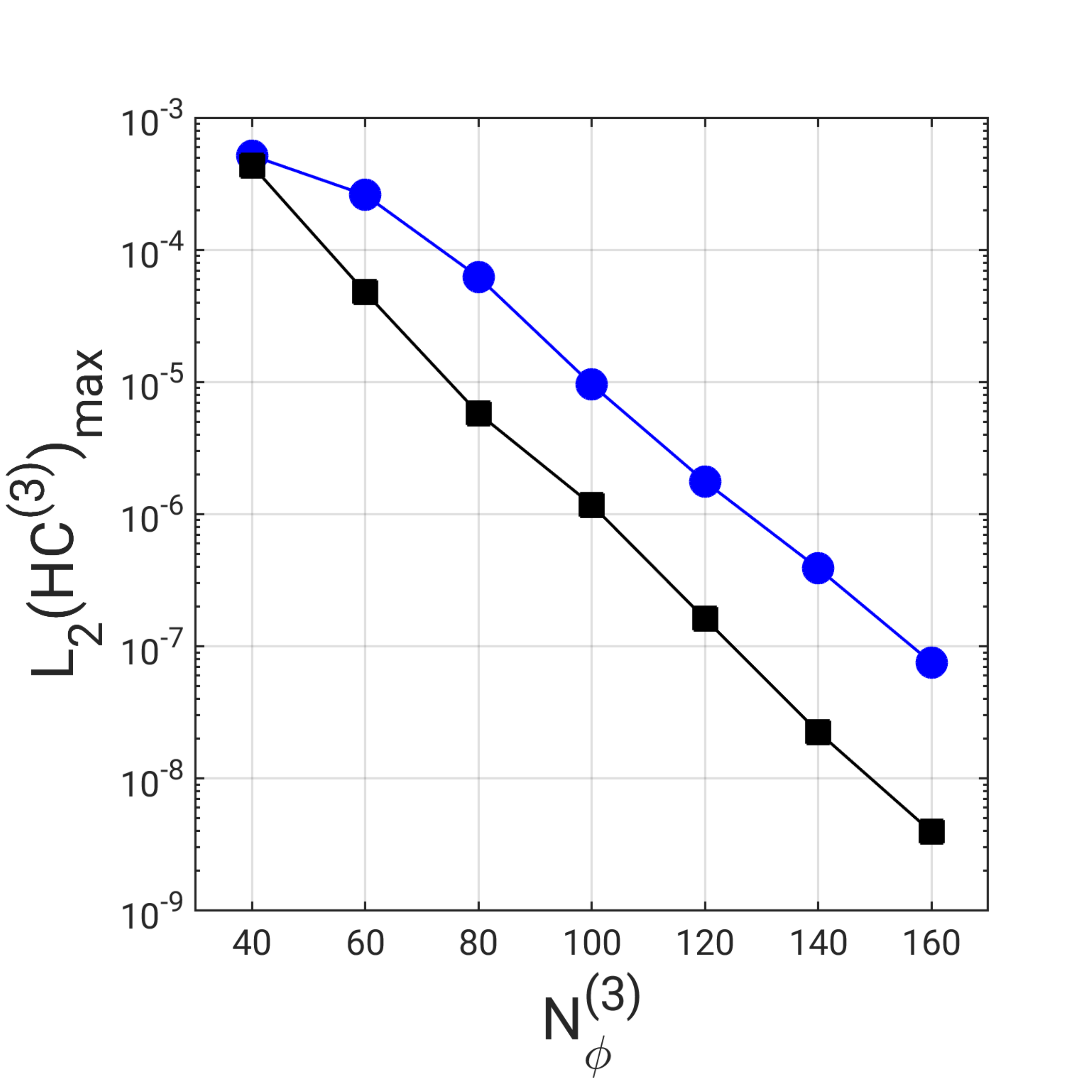}
	\caption{Decay {of the maximum value $L_2$-error at any time in the evolution,} associated with the Hamiltonian constraint in the first, second and third subdomains, from up to down. We have used two values of the map parameter, $L_0=5$ (blue circles) and $10$ (black boxes). $N_\phi^{(1)}, N_\phi^{(2)}{=N_\phi^{(1)} + 1}$ and $N_\phi^{(3)}{=1.3N_\phi^{(1)} + 1}$ are the truncation orders of the scalar field in the first, second, and third subdomains, respectively.}
\end{figure}

\begin{figure*}[htb]
	\centering
	\includegraphics[scale=0.27]{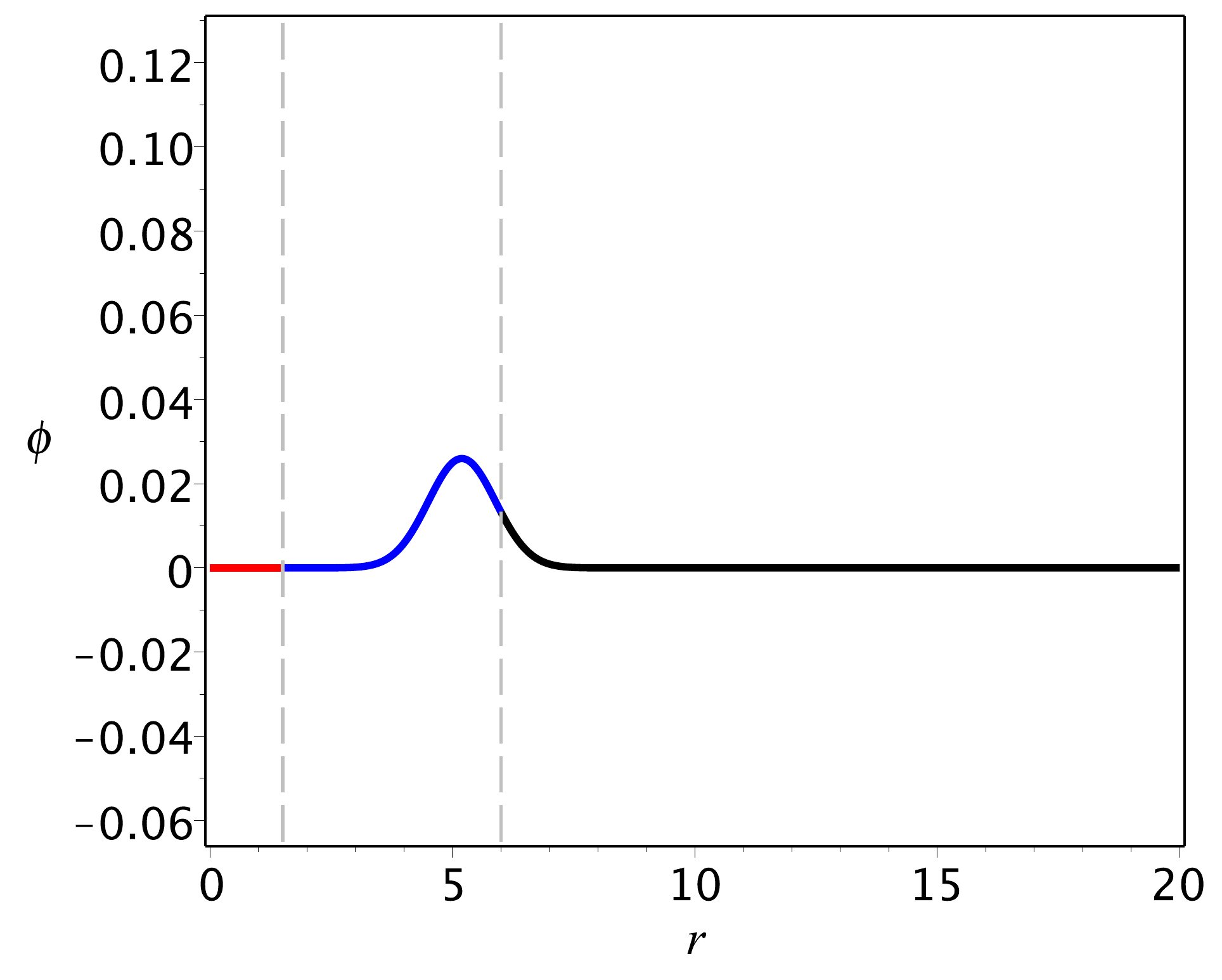}\includegraphics[scale=0.27]{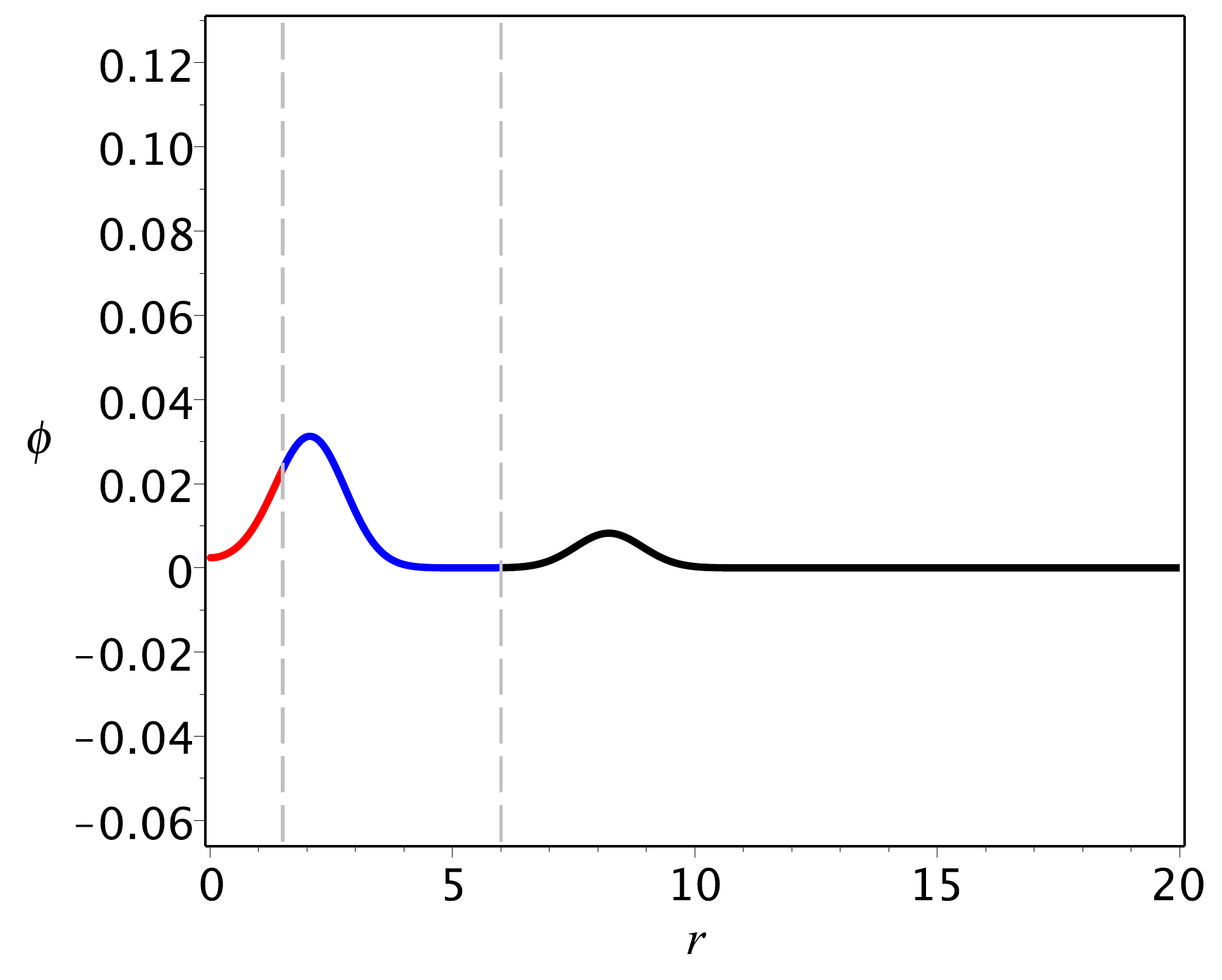}\includegraphics[scale=0.27]{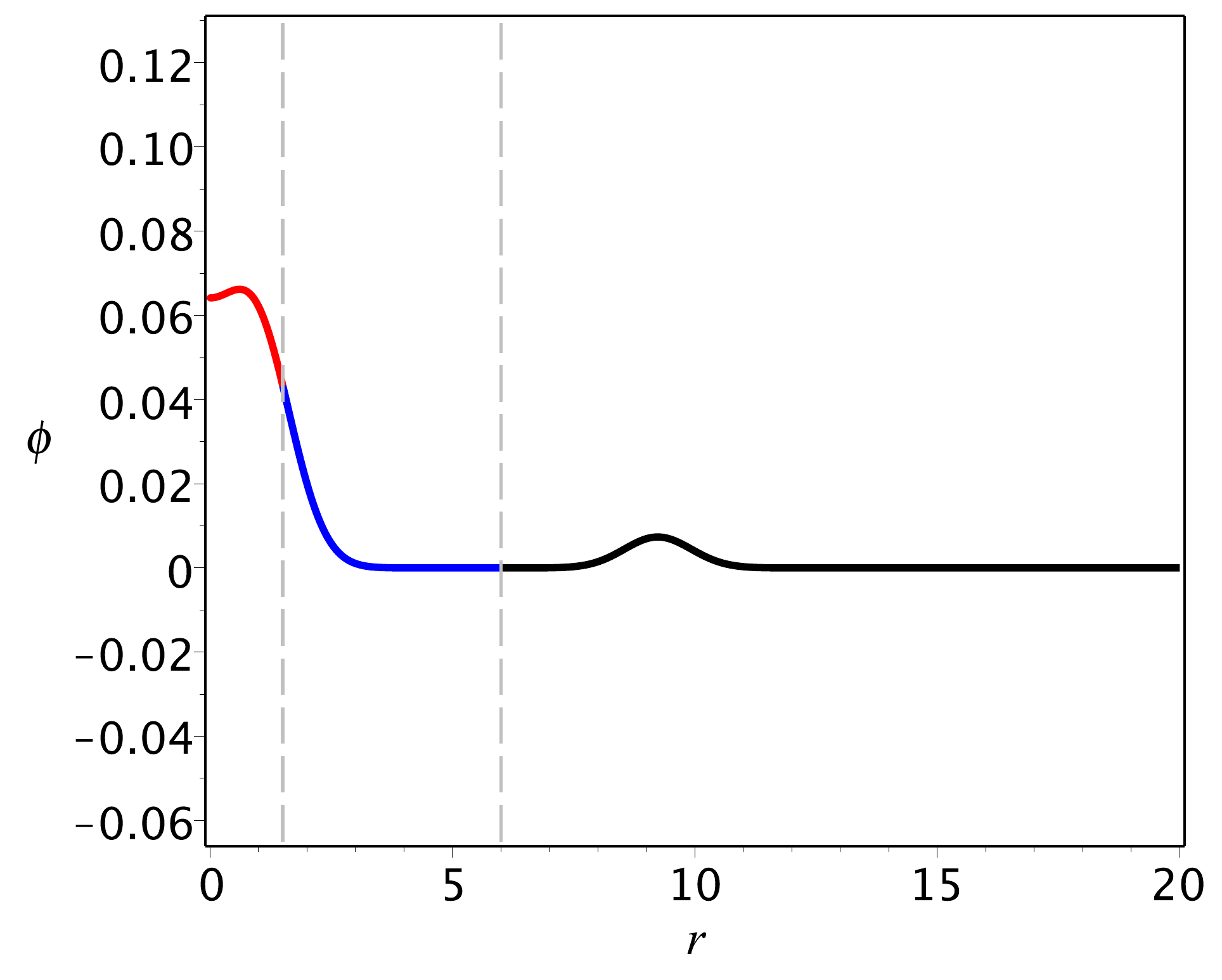}\\
	\includegraphics[scale=0.27]{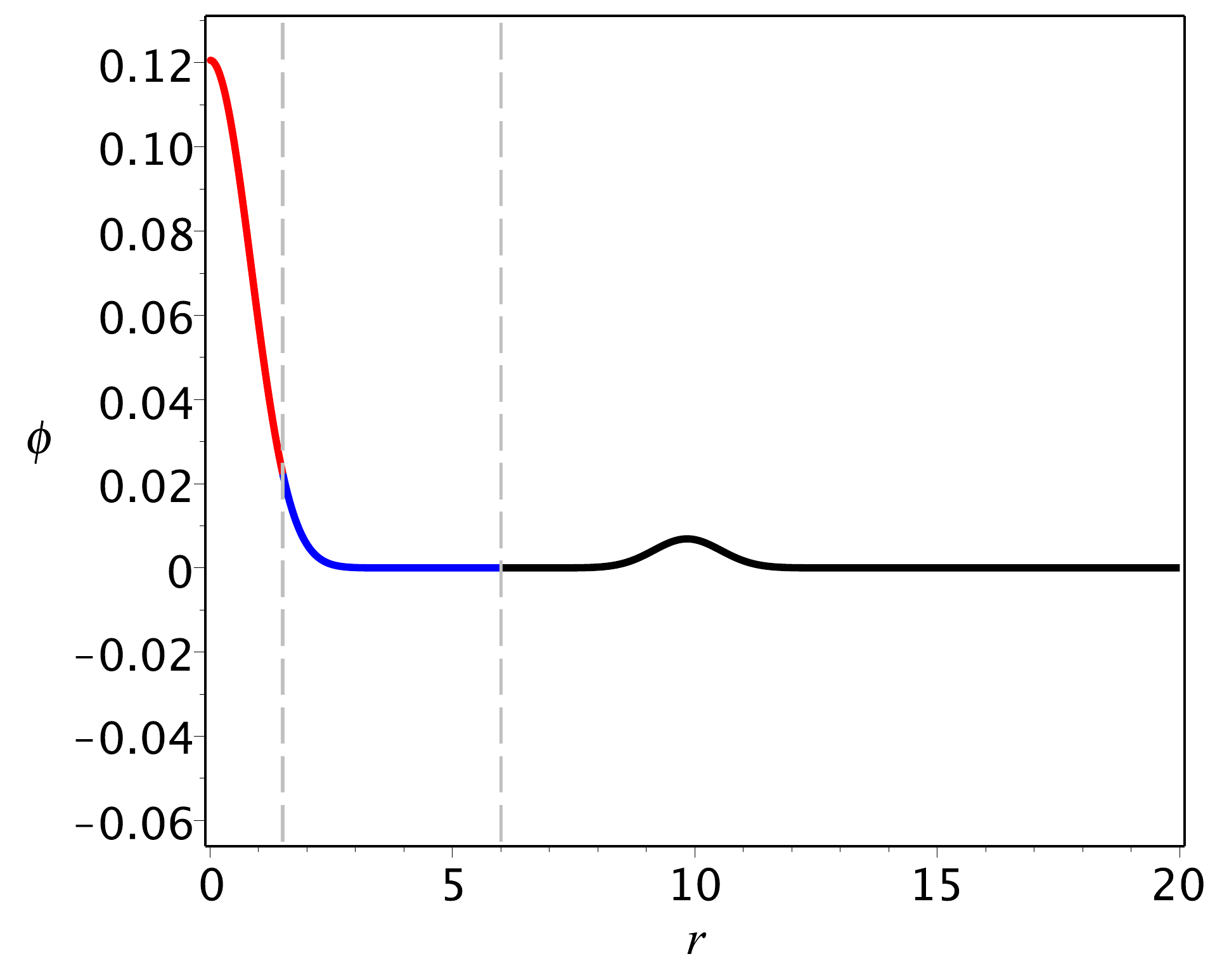}\includegraphics[scale=0.27]{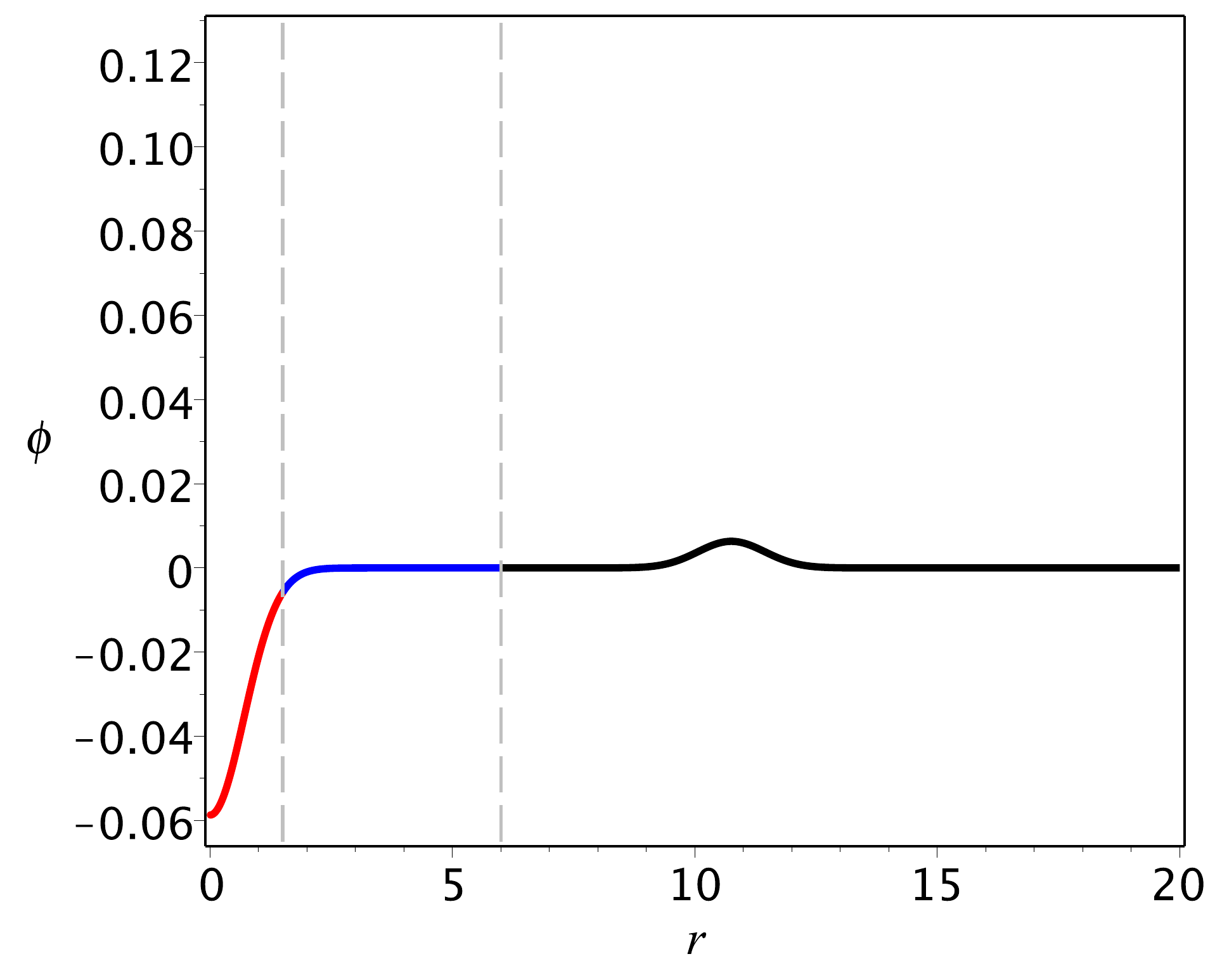}\includegraphics[scale=0.27]{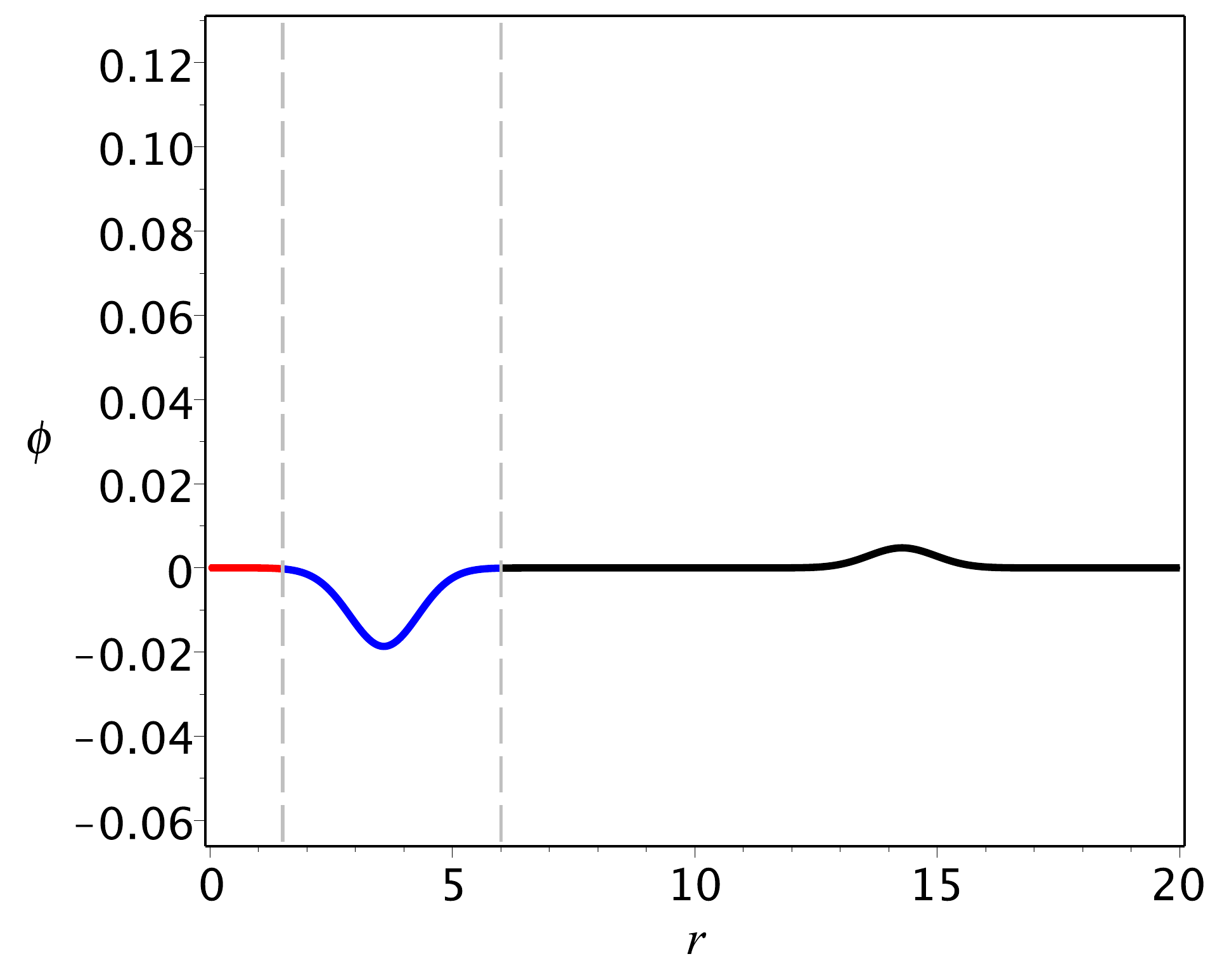}
	\caption{Snapshots of the scalar field calculated at $t=0,3$, and $4$ (left to right in top row), and at $t=4.6,5.5,$ and $10$ (left to right in bottom row). Here, the initial amplitude is $A_0=0.002$ and the pulse is centered initially at $r_0=5$ (cf. Eq. (\ref{eq51})). The line colors red, blue, and black indicate the first, second, and third subdomains, respectively; the vertical dashed lines correspond to each interface.}
\end{figure*}


We calculate the above $L_2$-error with two and three subdomains using the same truncation orders of Fig. 2 and 3. In Fig. 5, we show the exponential decay of the maximum {value of the $L_2$-error at any time in the evolution,} in the first and second subdomains for $L_0=5$ (blue circles)  and $L_0=10$ (black boxes). The effect of increasing the map parameter produces an increase of the maximum error in the first subdomain ($\mathcal{D}_1:\,0 \leq r \leq L_0$), but still, the maximum error decay is exponential. Notice that increasing the subdomain's size means a lower numerical resolution and larger errors if we fix the truncation order. For instance, for $N_\phi^{(1)}=120$, we obtain $L_2(HC^{(1)})_{\mathrm{max}} \sim \mathcal{O}(10^{-7}),  \mathcal{O}(10^{-5})$ for $L_0=5$, and $10$, respectively. These results are very satisfactory, but the most important feature is the exponential decay of the maximum error that guarantees a fast code convergence. On the other hand, the opposite occurs for the second subdomain; that is, the increase of $L_0$ improves the error decay, as shown in the lower panel of Fig. 5. 

In the case of three subdomains, we observed an improvement of the maximum $L_2$-error associated with the Hamiltonian constraint in each subdomain, as shown in Fig. 6. The error decay exponentially in all three subdomains for $L_0=5$ and $L_0=10$. In particular, in the first subdomain, the maximum error saturation of order $10^{-11}$ is achieved for $N^{(1)}_\phi \geq 120$. {Eventually, the error saturates about the same order with $L_0=10$ after increasing the truncation orders.}   

{In the above numerical experiments, the choice of the map parameter plays a relevant role in favoring exponential convergence of the $L_2$-errors associated with the Hamiltonian constraint. We have tested other smaller values, $L_0=2,3$, but the results were not satisfactory. In a sense, $L_0$ organizes the subdomains' size and the corresponding collocation points. Then, smaller map parameters produced dense inner subdomains with small errors and large outer subdomains. In this situation, the convergence is generally poor.}

Next, we reproduce the snapshots of the scalar field's evolution, starting with the amplitude $A_0=0.001$, the same adopted in Ref. \cite{alcubierre}. As stated in this reference, this amplitude is large but not enough to form an apparent horizon. We have considered the evolution produced by a three subdomain code with map paramter $L_0=3$, $N_\phi^{(1)} = 80$, $N_\phi^{(2)} = 81$, and $N_\phi^{(3)} = 88$ with the snapshots displayed in Fig. 7 corresponding to the instants $t=0,3,4,4.6,5.5,10$. In the panels, we included the location of the subdomains with the red, blue, and black lines correspond to the first, second, and third subdomains, respectively.

The scalar field pulse separates into two parts traveling in opposite directions. The ingoing pulse implodes at the origin, reaching to a maximum amplitude of $0.12$ while the outgoing component is located at $r \approx 10$ in agreement with the simulation of Ref. \cite{alcubierre}. The initially ingoing pulse changes its sign after imploding through the origin and becomes an outgoing pulse, as shown in the last panel evaluated at $t=10$.

A valuable test for the GC domain decomposition algorithm is to observe an apparent horizon's formation and estimate the mass inside it.  The apparent horizon forms when the expansion of outgoing null rays, $\Theta^+$, vanishes \cite{baumagarte}. With the coordinate system under consideration, we have that $\Theta^+=0$ implies 
\begin{equation}
\mathrm{e}^{-2 \mu(t_{\mathrm{AH}},r_{\mathrm{AH}})} \rightarrow 0, \label{eq58}
\end{equation}

\begin{figure}
	\includegraphics[width=7.cm,height=6cm]{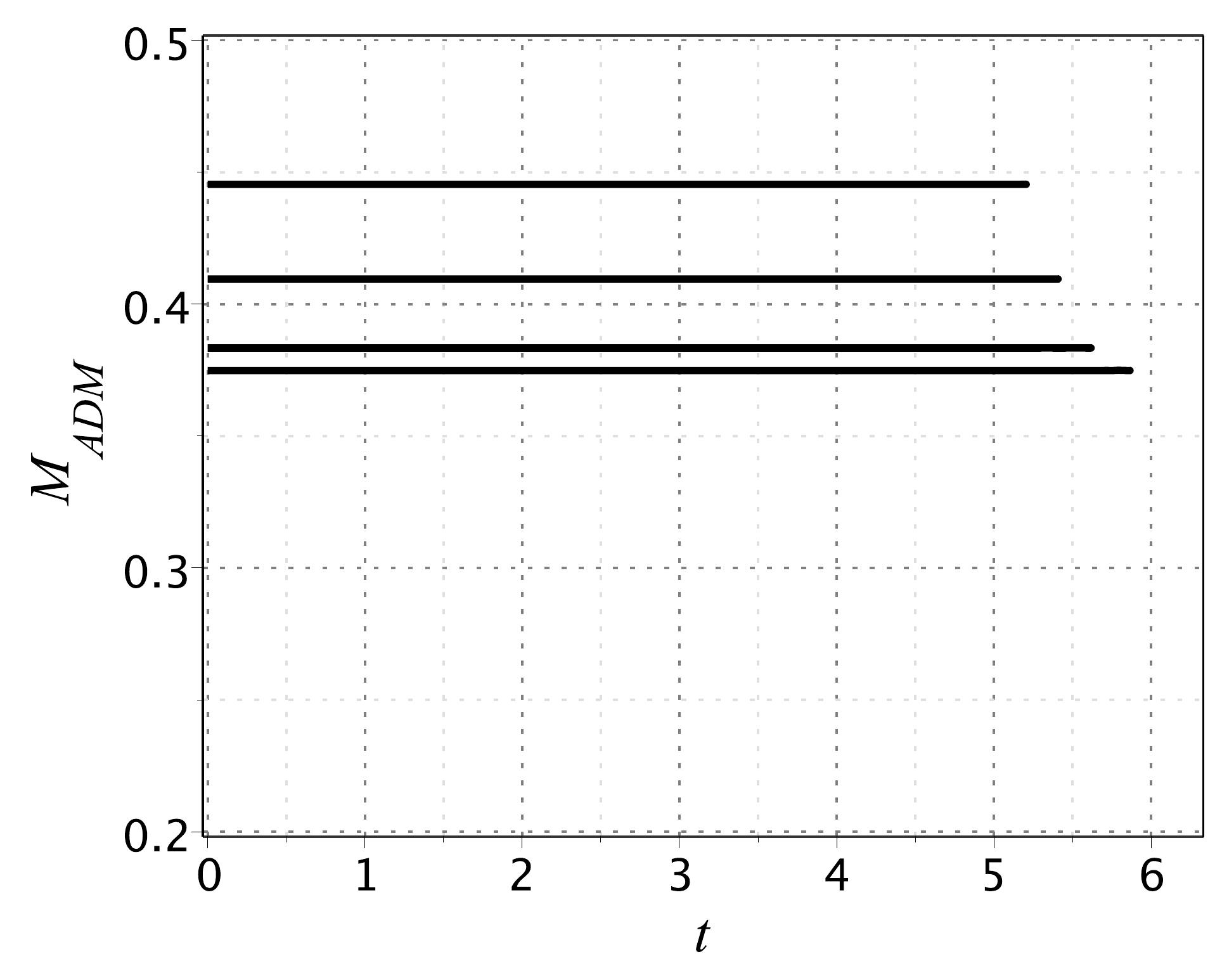}
	\caption{ADM masses corresponding to the initial amplitudes $A_0=0.0082, 0.0083, 0.0086$, and $0.009$, from bottom to top. In all cases a black hole is formed.}
\end{figure}

\noindent where $t_{\mathrm{AH}}$ is the time of the apparent horizon formation located at $r_{\mathrm{AH}}$. From the above condition. we  calculate the mass at the apparent horizon directly from Eq. (\ref{eq8}) as
\begin{equation}
m_{\mathrm{AH}} = \frac{r_{\mathrm{AH}}}{2}, \label{eq59}
\end{equation}

\noindent after determining $r_{\mathrm{AH}}$. However, Eq. (\ref{eq59}) implies that the metric function $\mu(t_{\mathrm{AH}},r_{\mathrm{AH}})$ diverges, and as a consequence, spoiling the numerical integration. 

We have performed simulations starting with the initial data (\ref{eq53}) with $r_0=5.0$. We have previously mentioned that the apparent horizon forms if $A_0  \gtrsim 0.0082$, which represents a discrepancy with Ref. \cite{alcubierre}. {In that work Alcubierre finds that a black hole is formed for $A_0$ between $0.001$ and $0.002$}. We present some ADM mass plots versus time in Fig. 8 for those solutions ending up forming an apparent horizon, which corresponds to the initial amplitudes $A_0=0.0082, 0.0083, 0.0086$, and $0.009$. The ADM masses remain constant until the formation of an apparent horizon when the integration breaks up. As expected, the dynamics takes a longer time for smaller initial amplitudes. We have performed the numerical simulations using a three-subdomains code with $N_\phi^{(1)}=90$ and map parameter $L_0=2.0$. 
%

\begin{figure}[htb]
	\includegraphics[width=7.cm,height=6cm]{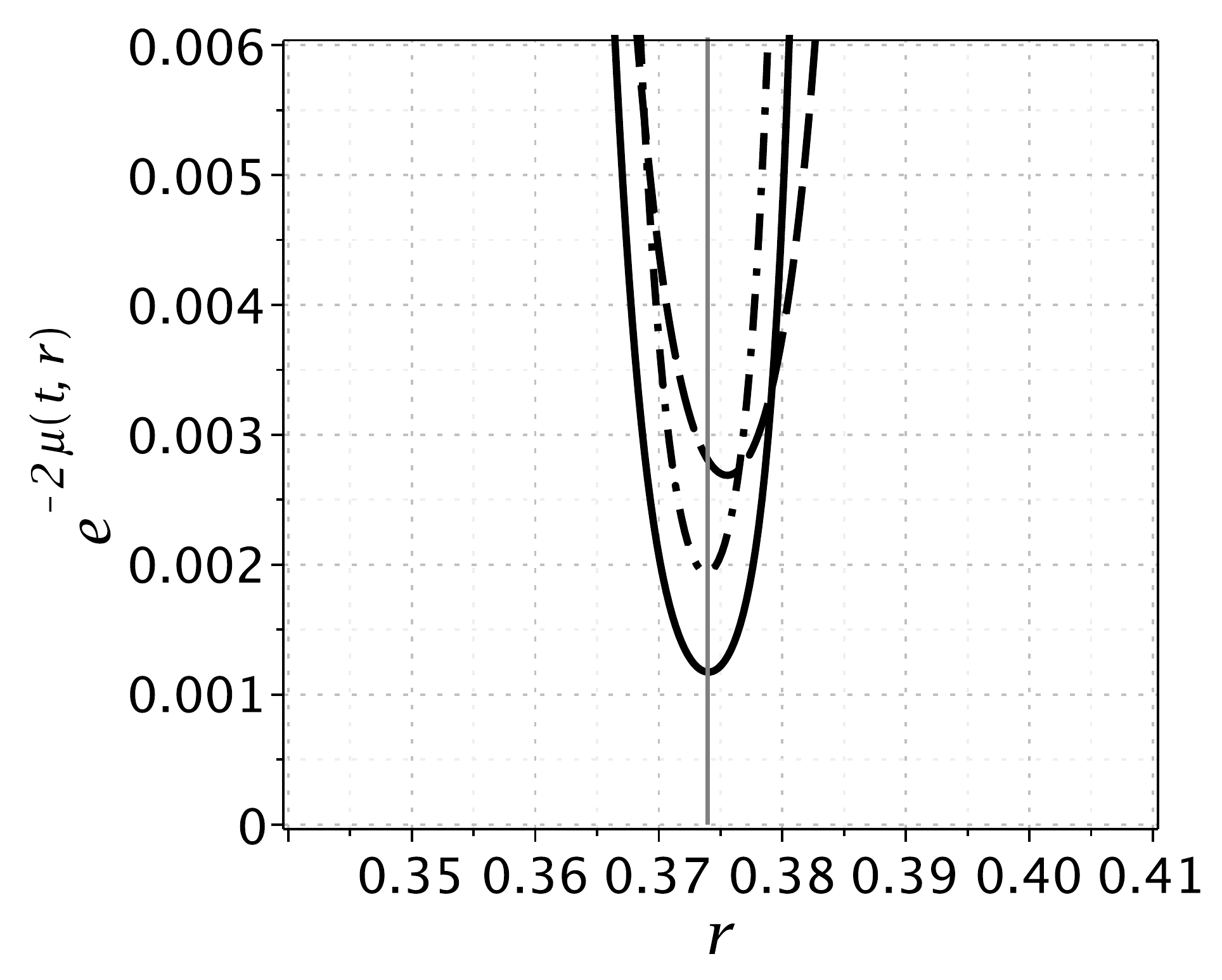}
	\caption{Plots of $\mathrm{e}^{-2 \mu(t,r)}$ when the apparent horizon forms for distinct resolutions with a three subdomains code. The dashed, dot-dashed and continuous lines correspond to $N^{(1)}_\phi=120, 140$ and $160$, respectively. {In all three cases the time is about 2.514 3238.}}
\end{figure}

\begin{figure}[htb]
	\includegraphics[width=7.cm,height=6cm]{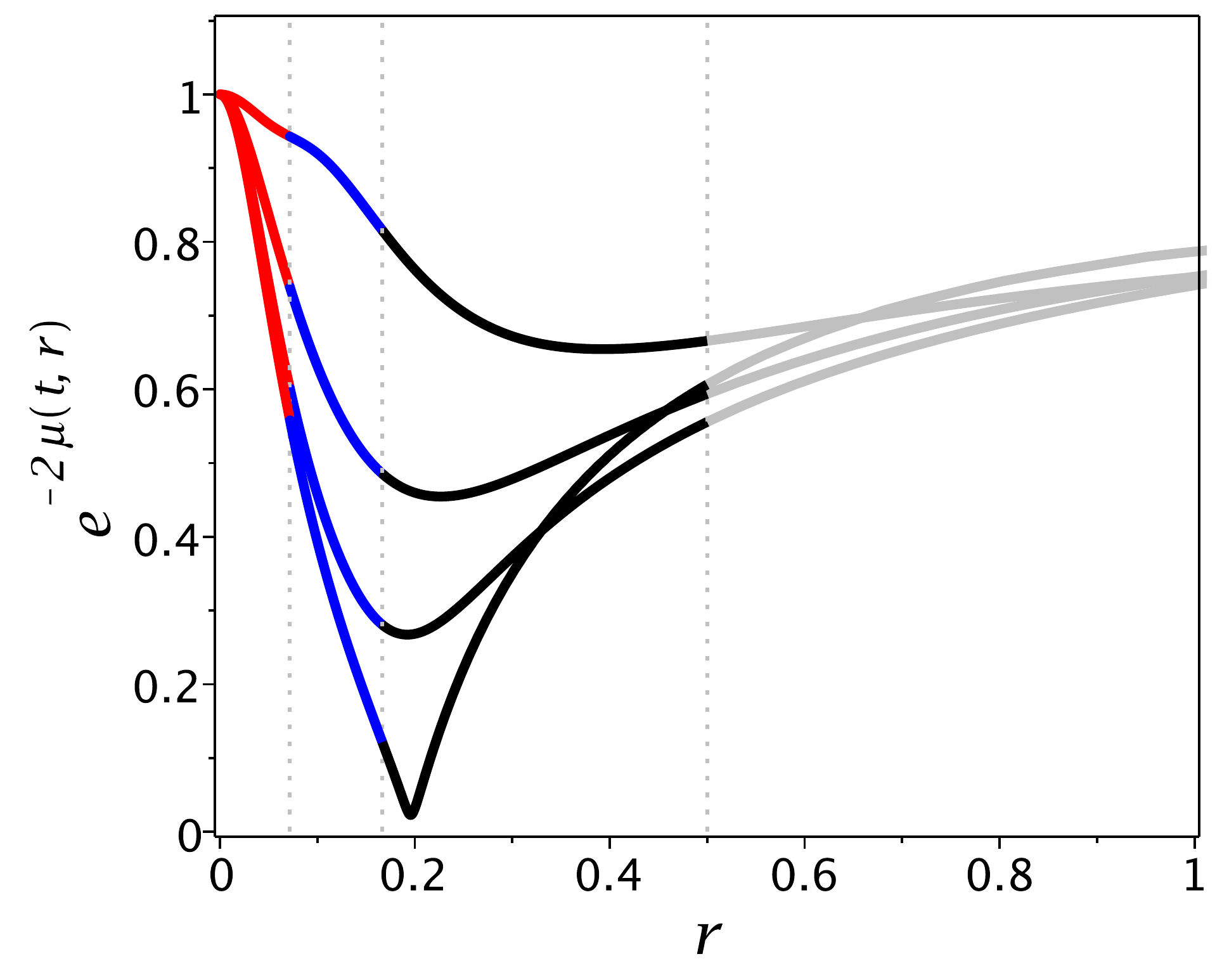}\\
	{\center \vspace{-0.5cm} (a)}\\
	\includegraphics[width=7.5cm,height=6cm]{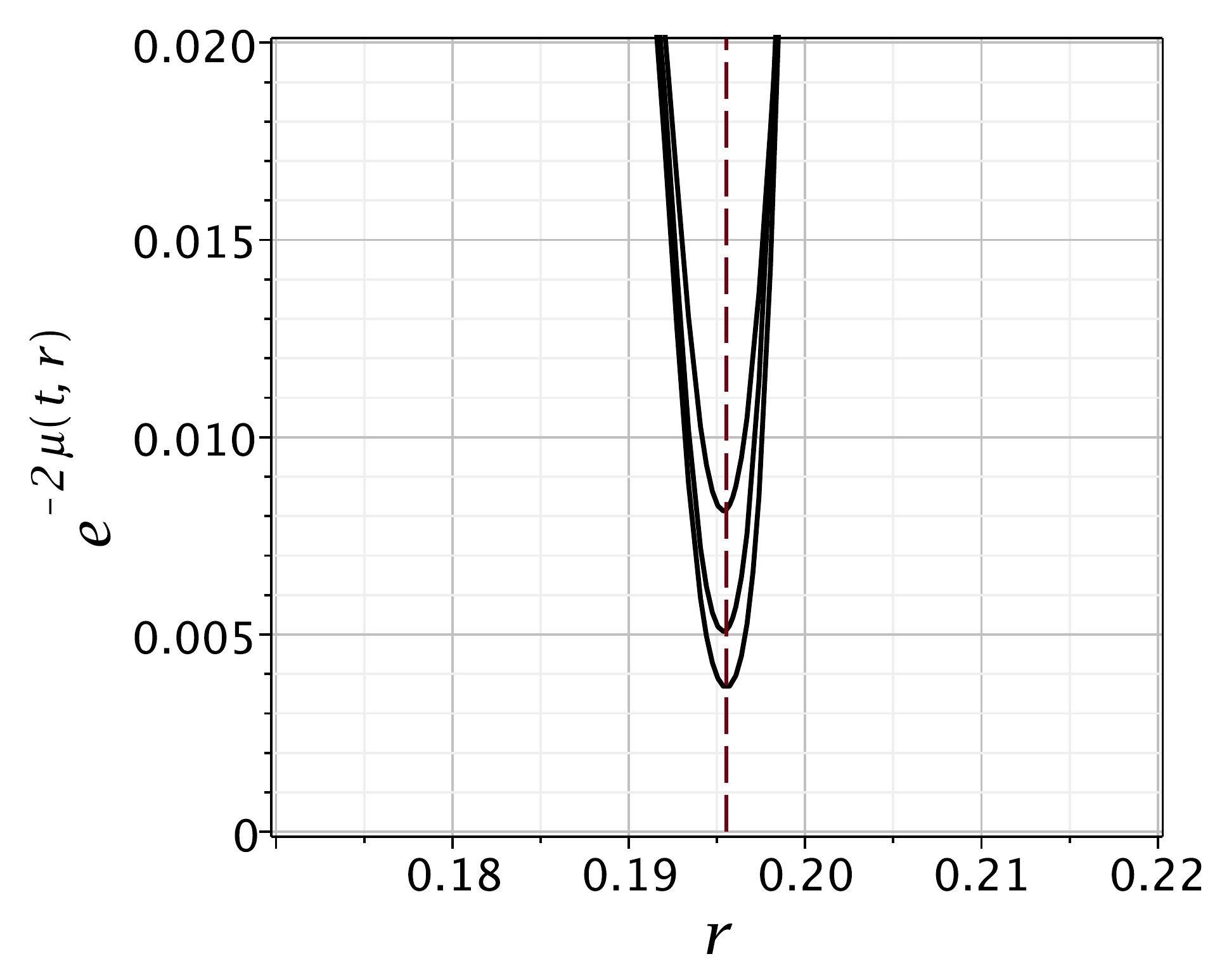}\\
	{\center \vspace{-0.5cm} (b)}
	\caption{(a) Plots of $\mathrm{e}^{-2 \mu(t,r)}$ showing the approach to the apparent horizon. The curves are evaluated at $\approx 2.802, 2.876,2.892,2.896$. (b) Fine structure of the approach to the apparent horizon prior the integration is halted. The curves are calculated at $t \approx 2.895\,632\,50, 2.895\,632\,9, 2.895\,634\,40$ from up to down. Here $A_0=0.088$}
\end{figure}




We proceed with the determination of the mass at the apparent horizon. To this aim, we have implemented a four subdomain code and set $r_0=2.0$ in the initial data (\ref{eq53}) to reduce the integration time. According to Eq. (59), it is necessary to determine the location and the instant the apparent horizon forms, or $(t_{AH},r_{AH})$. Then, we need to obtain the minimum value of $\mathrm{e}^{-2 \mu(t,r)}$ at the end of the integration, where $\mu(t,r)$ is expressed by Eq. (21). To provide a qualitative convergence test to determine the apparent horizon location, we have considered the previous code with three subdomains but with increasing resolutions as $N^{(1)}_\phi=120,140,160$. We fix the initial amplitude $A_0=0.096$ (for $A_0 \gtrsim 0.087$ an apparent horizon forms) and map parameter $L_0=2$. 
{We also evolved the dynamical system with the adaptive Cash-Karp integrator \cite{ck} with the following parameters: $h_{\mathrm{min}}=10^{-15}$ and  $h_{\mathrm{max}}=10^{-4}$, as the minimum and maximum time steps, respectively. The tolerance is set to $10^{-9}$. With these parameters we have achieved exponential convergence.} The maximum time duration of the simulations was about 5 hours using a notebook with a $9^{th}$ generation core $i7$ processor. We present the results in Fig. 9 with the snapshots of $\mathrm{e}^{-2 \mu(t,r)}$ evaluated at the end of the integrations with $N^{(1)}_\phi=120, 140$ and, $160$ represented by dashed, dot-dashed and continuous lines, respectively. {In all three cases, the time is about 2.514 3238. The time difference between each curve is tiny. By increasing the resolution, we have reached a minimum $\mathrm{e}^{-2 \mu(t_{AH},r_{AH})}$ of about $10^ {-3}$ with $r_{AH} \approx 0.3738$. We have obtained the same result with higher resolution or more subdomains, suggesting we have reached the threshold of the apparent horizon whenever $\mathrm{e}^{-2 \mu(t_{AH},r_{AH})} \approx 10^ {-3}$. We believe that a possible explanation is that the metric function $\mu(t,r)$ is described globally with the radial dependence given by smooth basis functions. On the other hand, we have noticed that the modal coefficients tend to have high absolute values in approaching the apparent horizon formation. At the same time, the profile of the metric function $\mu(t,r)$ becomes frozen and smooth inside the region $r \leq r_{AH}$, as expected, but implying finite modes inside the horizon. Nevertheless, Fig. 9 suggests a convergence.}

In the next numerical experiments, we have implemented a four subdomains code $x^{(1)}=-0.75,x^{(2)}=-0.5,x^{(3)}=0$ in the computational auxiliary domain $-1 \leq x \leq 1$. In the physical domain, these interfaces are located at: 

\begin{equation*}
r^{(1)}=\frac{L_0}{7},\;r^{(2)}=\frac{L_0}{3},\;r^{(3)}=L_0.
\end{equation*}

\noindent  Therefore, most collocation points are inside the region $0 \leq r \leq L_0$. Depending on the map parameter $L_0$, we can concentrate the colocation points near the origin where the apparent horizon forms. For instance, considering $A_0=0.088$ the ADM mass is about $0.361$ that corresponds to a $r_{AH} \approx 0.72$ \textit{if} all mass is trapped inside the horizon, that naturally is not the case. Thus, the apparent horizon will necessarily form at $r_{AH} < 0.72$, implying that a reasonable choice would be $L_0=0.5$.

\begin{figure*}[htb]
	\includegraphics[width=7.cm,height=6cm]{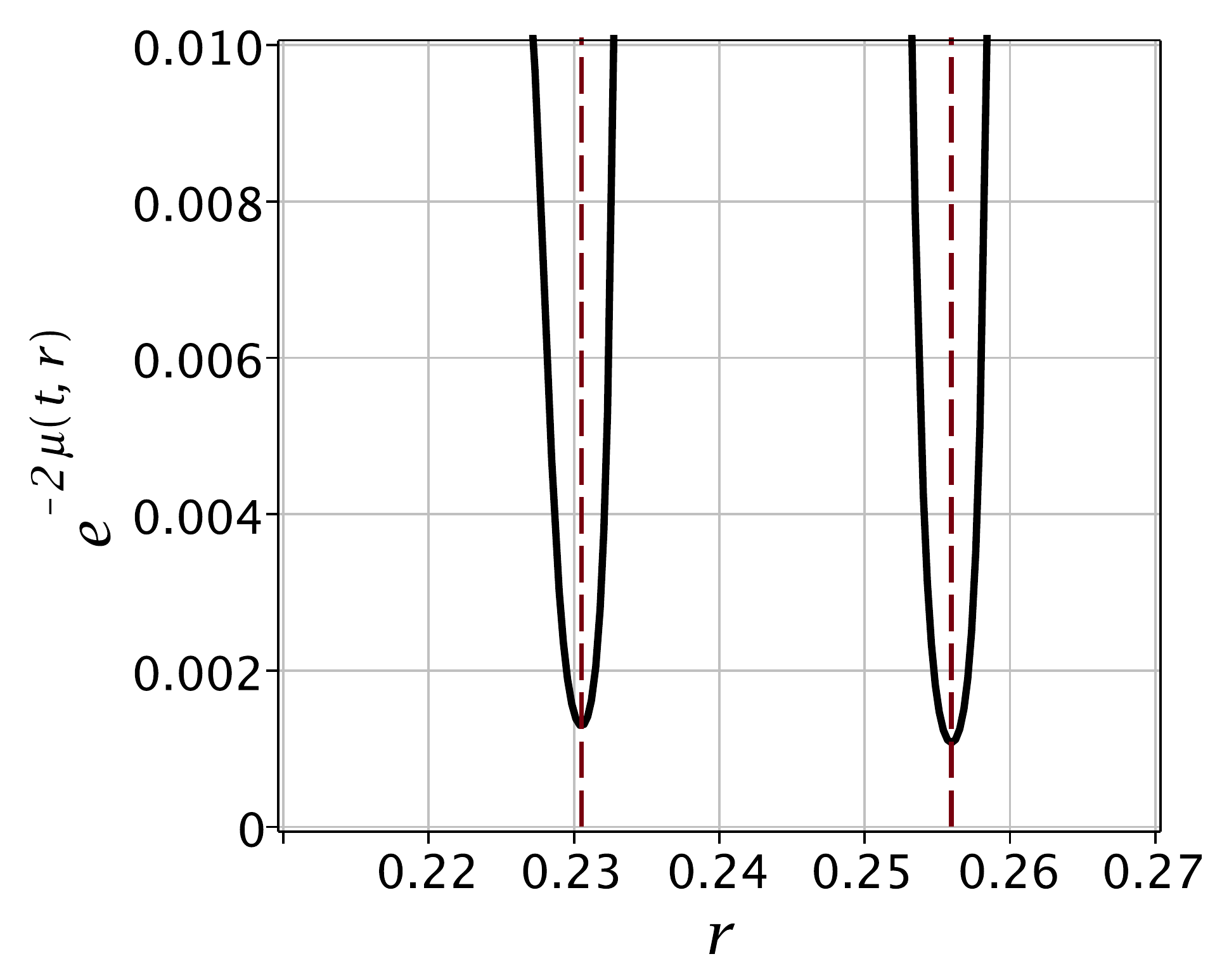}\includegraphics[width=7.cm,height=6cm]{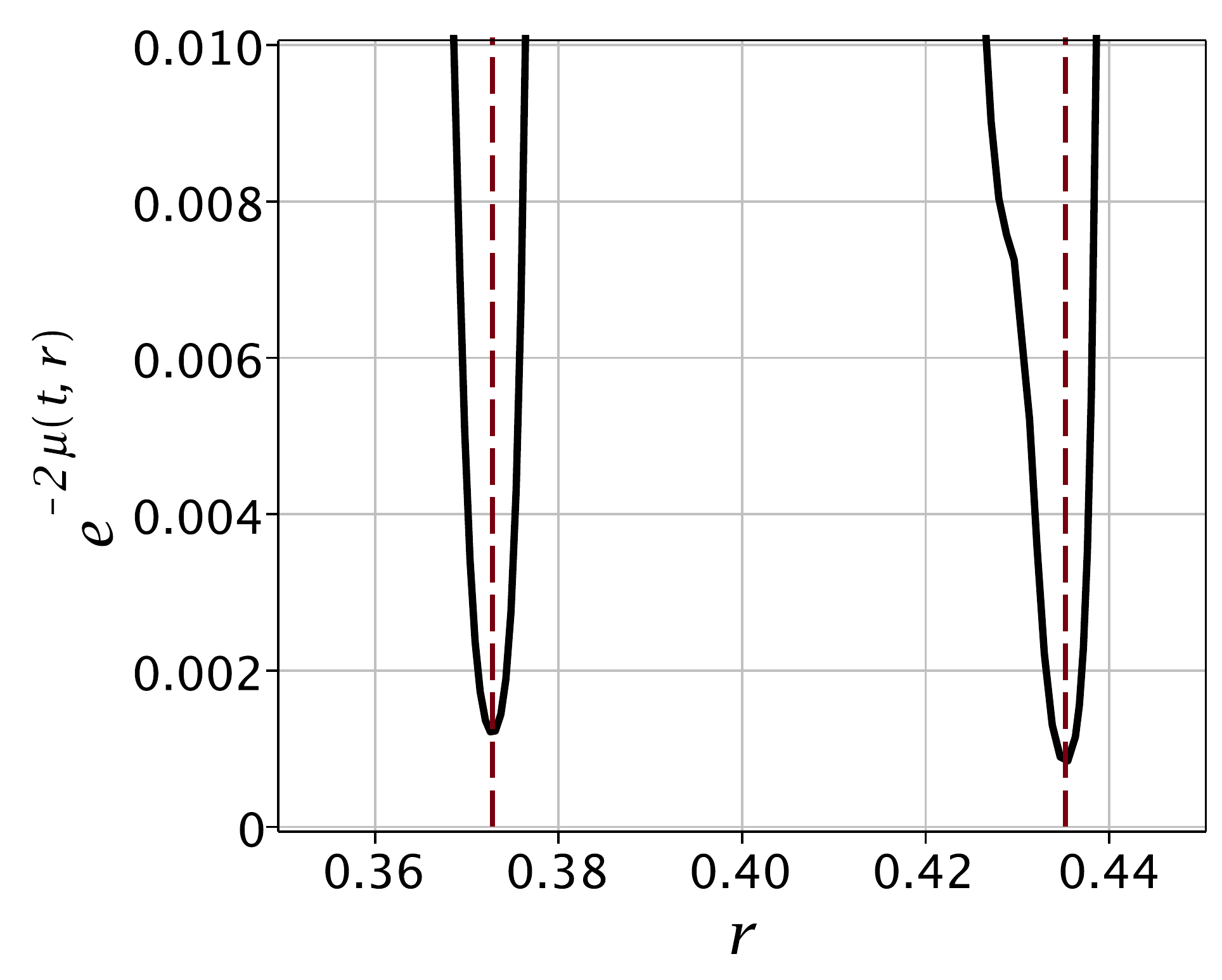}
	\caption{Snapshots of $\mathrm{e}^{-2 \mu(t,r)}$ evaluated at the apparent horizon formation for $A_0=0.089,0.090,0.096$, and $0.10$, from left to right. The dashed lines indicate the location of the apparent horizon radius in each case. The masses evaluated at the apparent horizon are $M_{AH} \simeq 0.1153,\,0.1280,\,0.1864,\,0.2252$} 
\end{figure*}

We present in Fig. 10 snapshots of $\mathrm{e}^{-2 \mu(t,r)}$ evaluated in several instants showing the approach to the formation of an apparent horizon. Here, the code has $N_\phi^{(1)}=100$. We have indicated the subdomains' boundaries by dotted vertical lines, and the colors red, blue, black, and gray correspond to the first, second, third, and fourth subdomains. In Fig. 10(a), the plots from up to down are calculated at $t \approx 2.802, 2.876,2.892,2.896$. It becomes evident the fast approach to the apparent horizon formation. The detailed approach to the apparent horizon is shown in Fig. 10(b), where the three graphs of $\mathrm{e}^{-2 \mu(t,r)}$ located in the third subdomain approach to $r=r_{AH}$ whose minimum value is of the order of $\mathcal{O}(10^{-3})$. Notice that this minimum value takes place at $r_{AH}  \simeq 0.195\,53$ that corresponds to the apparent horizon mass $M_{AH} \simeq 0.097\,8$. It is important to mention that the adaptive time step integrator was crucial to following the scalar field dynamics until the apparent horizon formed. The graphs of Fig. 10(b) correspond to $t \approx 2.895\,632\,50, 2.895\,632\,9, 2.895\,634\,40$.


In the next numerical experiments, we have considered the initial amplitudes $A_0=0.089, 0.090, 0.096$ and $0.10$ whose corresponding ADM masses are $M_{ADM} \simeq 0.368\,458, 0.375\,726,0.420\,091$, and $0.450\,307$, respectively.  We performed the numerical simulations using a four subdomain code with better resolution, namely $N_\phi^{(1)}=130$. Then, in Fig. 11, we present the plots of $\mathrm{e}^{-2\mu(t,r)}$ evaluated at the end of the integration at the instants $t \approx 2.836, 2.768, 2.514,2.407$. The location of the apparent horizon in each case is determined by the minimum of  $\mathrm{e}^{-2\mu(t,r)}$, $r=r_{AH}$ indicated by the dashed lines. In the left panels, plots correspond to $A_0=0.089$ and $0.09$, whereas in the right panels $A_0=0.096$ and $0.10$; the apparent horizon masses are $M_{AH} \simeq 0.1153,\,0.1280,\,0.1864,\,0.2252$. Therefore, these results mean that the GC domain decomposition algorithm is effective is determining the apparent horizon masses. 

\section{Final comments}

The present work is the second of a systematic development of the multidomain or domain decomposition technique connected with the Galerkin-Collocation method and applied to situations of interest in numerical relativity.  In the continuity of Ref. \cite{alcoforado}, we have considered the self-gravitating and spherically symmetric scalar field in the $3+1$ formulation.

We have used the same numerical scheme of Ref. \cite{alcoforado}: we compactified and divided the spatial domain $\mathcal{D}: 0 \leq r < \infty$ into several subdomains.  An intermediate computational domain $-1 \leq x \leq 1$ is introduced using an algebraic map given by (\ref{eq24}), and with linear mappings, we define several subdomains labeled by $-1 \leq \xi^{(l)} \leq 1,\;\;l=1,2,..,n$ as shown by Fig. 1. This scheme is the backbone for implementing the algorithm.

The basis functions for the scalar field and the metric functions have a distinctive feature of satisfying the boundary conditions. In particular, the basis functions provided the exact cancellation of the $1/r$ terms near the origin, therefore eliminating a potential source of numerical errors. It is important to emphasize that the evident advantage of canceling the $1/r$ terms near the origin is that these terms, if not canceled, would quite likely lead to an unstable time evolution.  

We have adopted the patching method \cite{canuto_88} for joining the subdomain solutions through the transmission conditions in the subdomains' interfaces. It represents the most straightforward version applied to a hyperbolic system since, for these categories of problems, the transmission conditions are not uniquely defined \cite{kopriva_86,kopriva_89}. 

We have presented numerical tests to validate the code, starting with initial data whose amplitude is close to the minimum necessary for forming an apparent horizon. We adopted two error measures: the relative deviation of the ADM mass and the $L_2$-errors associated with the Hamiltonian constraint in each subdomain. In all cases, we noticed that these error measures exhibit a spectral convergence; in other words, they decay exponentially with the increase of the truncation orders. The exponential convergence also depends on the other factors like the map parameter $L_0$ and the interfaces defining the subdomains size. We remind that the Hamiltonian constraint is used to determine the initial data $\mu(t_0,r)$. 

It was possible to obtain the mass inside the apparent horizon by searching the minimum of $\mathrm{e}^{-2\,\mu(t,r)}$ when the numerical integration diverges, which signalizes the formation of an apparent horizon. The introduction of multiple subdomains allowed us to concentrate more collocation points near the region where the apparent horizon forms after the scalar field hit the origin. For the case of four subdomains, with 130 collocation points in each subdomain, we have obtained $\mathrm{e}^{-2\,\mu(t,r)} \sim \mathcal{O}(10^{-3})$, despiting being described by an analytical function. The increase of collocation points can result in smaller levels of $\mathrm{e}^{-2\,\mu(t,r)}$ at the apparent horizon. 

{We want to extend our alternative approach, within the family of spectral methods, for higher dimensions and to the standard formulation in Numerical Relativity. 
Currently it is not possible for us do some comparison
with more sophisticated and complex codes.
It is expected that for a 2+1-D and 3+1-D cases the matrices might be fairly sparse if there are a large number of domains, so sparse-matrix methods might be able to somewhat alleviate the cost.\footnote{We thank to one of the Referees for this quote.}} 

Finally, we would point out the next steps of the present investigation. We intend to continue developing the GCDD for the dynamics in spherically symmetric spacetimes considering the field equations in the BSSN formulation \cite{nok87,sn95,bs99, brown} in which we can find relevant works \cite{akbarian_choptuik,sorkin_choptuik,alcubierre_guzman,montero_baumgarte,montero_carrion_13,alcubierre_mendez}. For the case of a non-spherically symmetric spacetime, we are developing a GCDD algorithm for the Bondi problem \cite{alcoforado2}.

\begin{acknowledgments}
M. A. acknowledges the financial support of the Brazilian agency Coordena\c c\~ao de Aperfei\c coamento de Pessoal de N\'ivel Superior (CAPES). W. O. B. thanks to the Departamento de Apoio \`a Produ\c cao Cient\'\i fica e Tecnol\'ogica (DEPESQ) for the financial support, also to the Departamento de F\'\i sica Te\'orica for the hospitality, both at the Universidade do Estado do Rio de Janeiro. H. P. O. thanks Conselho Nacional de Desenvolvimento Cient\'ifico e Tecnol\'ogico (CNPq). 
\end{acknowledgments}


\begin{thebibliography}{99}
	
\bibitem{alcoforado} M. Alcoforado, W. Barreto and H. P. de Oliveira, Gen. Rel. Grav. 53, 42 (2021).

\bibitem{oliveira_14} H. P. de Oliveira and E. L. Rodrigues, Phys. Rev. D 90, 124027 (2014).

\bibitem{barreto_18} W. Barreto, P. C. M. Clemente, H. P. de Oliveira and B. Rdriguez-Mueller, Phys. Rev. D 97, 104035 (2018).

\bibitem{barreto_18_2} W. Barreto, P. C. M. Clemente, H. P. de Oliveira and B. Rodriguez-Mueller, Gen. Rel. Grav. 50, 71 (2018).

\bibitem{barreto_19} W. Barreto, J. A. Crespo, H. P. de Oliveira and E. L. Rodrigues, Class. Quant. Grav. 36, 215011 (2019). 

\bibitem{crespo_19} J. A. Crespo, H. P. de Oliveira and J. Winicour, Phys. Rev. D 100, 104017 (2019).

\bibitem{nok87} T. Nakamura, K. Oohara, and Y. Kojima, Prog. Theor. Phys. Suppl. 90, 1 (1987). 

\bibitem{sn95} M. Shibata and T. Nakamura, Phys. Rev. D 52, 5428 (1995).

\bibitem{bs99} T. W. Baumgarte and S. L. Shapiro, Phys. Rev. D 59, 024007 (1999).

\bibitem{brown} J. Brown, Phys. Rev. D 79, 104029 (2009).

\bibitem{deppe} N. Deppe, L. E. Kidder, M. A. Scheel, and S. A. Teukolsky, Phys. Rev. D 99, 024018 (2019).

\bibitem{canuto_88} C. Canuto, A. Quarteroni, M. Y. Hussaini and T. A. Zang, \textit{Spectral Methods in Fluid Dynamics}, Springer-Verlag (1988).

\bibitem{canuto_new} C. Canuto, A. Quarteroni, M. Y. Hussaini and T. A. Zang, \textit{Spectral Methods - Evolution to Complex Geometries and Applications to Fluid Dynamic}, Springer-Verlag (2007).

\bibitem{orszag_80} S. Orszag, J. Comp. Phys. 37, 70 (1980).

\bibitem{kopriva_86} D. A. Kopriva, Appl. Numer. Math. 2, 221 (1986).

\bibitem{kopriva_89} D. A. Kopriva, SIAM J. Sci. Sta. Comp. 10, 1 120 (1989).

\bibitem{faccioli_96} E. Faccioli, F. Maggio, A. Quarteroni and A. Tagliani, Geophysics 61, 1160 (1996).

\bibitem{bona} S. Bonazzola, E. Gourgoulhon, M. Salgado and J. A. Marck, Astron. Astrophys. 278, 421 (1993).

\bibitem{pfeifer} H. P. Pfeiffer, \textit{Initial data for black hole evolutions}, Ph.D. thesis (2003), arXiv:gr-qc/0510016.

\bibitem{ansorg} M. Ansorg, Class. Quantum Grav. 24 S1 (2007).

\bibitem{spec} Spectral Einstein Code, https://www.black-holes.org/code/SpEC.html

\bibitem{lorene} LORENE (Langage Objet pour la Relativit\'e Num\'erique), http://www.lorene.obspm.fr

\bibitem{kidder_00} L. E. Kidder, M. A. Scheel and S. A. Teukolsky, Phys. Rev. D 62, 084032 (2000).

\bibitem{szilagyi_09} B. Szil\'agyi, L. Lindblom and M. A. Scheel, Phys. Rev. D 80, 124010 (2009).

\bibitem{kidder1} L. E. Kidder, S. E. Field, F Foucart, E. Schnetter, S. A. Teukolsky, A. Bohn, N. Deppe, P. Diener, F. H\'ebert, J. Lippuner, J. Miller, C. D. Ott, M. A. Scheel, and T. Vicent, J. Comp. Phys. 335, 84 (2017).

\bibitem{kidder2} F. H\'ebert, L. E. Kidder and S. A. Teukolsky, Phys. Rev. D 98, 0444041 (2018).

{\bibitem{hw08} J. S. Hesthaven and T. Warburton, \textit{Nodal discontinuous {G}alerkin methods:
algorithms, analysis, and applications} (Springer, New Tork, London, 2008).} 

\bibitem{Hemberger_13} D. A. Hemberger, M. A. Scheel, L. E. Kidder, B. Szil\'agyi, G. Lovelace, N. W. Taylor, S. A. Teukolsky, Class.
Quantum Grav. 30, 115001 (2013).

\bibitem{sxs_col} M. Boyle et al., \textit{The SXS Collaboration catalog of binary black hole simulations}, arXiv: 1904.04831 (2019).

\bibitem{boyd} J. P. Boyd, \textit{Chebyshev and Fourier Spectral Methods} (Dover Publications, New York, 2001).

\bibitem{alcubierre_potential} M. Alcubierre, F. S. Guzman, T. Matos, D. Nunez, L. A. Urena-Lopez and P. Wierderhold, Class. Quant. Grav. 19, 5017 (2002).

\bibitem{choptuik} M. W. Choptuik, Phys. Rev. Letters 70, 9 (1993).

\bibitem{shapiro_teukolsky_80} S. L. Shapiro and S. A. Teukolsky, The Astrophys. J. 235, 199 (1980).

\bibitem{ADM} R. Arnowitt, S. Deser and C. W. Misner, \textit{Republication of: The dynamics of general relativity}. Gen Relativ Gravit 40, 1997 (2008).

\bibitem{misner} C. W. Misner and D. H. Sharp, Phys. Rev. 136, B571 (1964).

\bibitem{MTW} C. W. Misner, K. S. Thorne and J. A. Wheeler, \textit{Gravitation}, W. H. Freeman  (1973).

\bibitem{peyret} R. Peyret, \textit{Spectral methods for incompressible viscous flow}, Springer-Verlag, New York (2002).

\bibitem{fornberg} B. Fornberg, \textit{A practical guide to pseudospectral methods}, Cambridge University Press (1998).

\bibitem{wtichy} {W. Tichy, Phys. Rev. D \textbf{80}, 104034 (2009).}

\bibitem{alcubierre} M. Alcubierre, \textit{Introduction to 3+1 numerical relativity}, Oxford University Press (2008).

\bibitem{baumagarte} T. W. Baumgarte and S. L. Shapiro, \textit{Numerical Relativity}, Cambridge University Press (2010).

\bibitem{ck} J. R. Cash and A. H. Karp, \textit{Variable order Runge-Kutta method for initial value problems with rapidly varying right-hand sides}, ACM Transactions on Mathematical Software 16, 201 (1990).

\bibitem{akbarian_choptuik} A. Akbarian and M. W. Choptuik, Phys. Rev D {\bf 92}, 084037 (2015).

\bibitem{sorkin_choptuik} E. Sorkin and M. W. Choptuik, Gen. Rel. Grav. 42, 1239 (2010).

\bibitem{alcubierre_guzman} M. Alcubierre, M. Corichi, J. A. Gonzales, D. Nunez, B. Reiman and M. Salgado, Phys. Rev. D 72, 124018 (2005).
 
\bibitem{montero_baumgarte} P. J. Montero, T. W. Baumgarte, I. Cordero-Carrion and E. Mueller, Phys. Rev. D 87, 044026 (2013).

\bibitem{montero_carrion_13} P. J. Montero and I. Cordero-Carrion, J. Phys. Conf. Ser. 454, 012002 (2013).

\bibitem{alcubierre_mendez} M. Alcubierre and M. D. Mendez, Gen. Rel. Grav. 43, 2769 (2011).

\bibitem{alcoforado2} M. Alcoforado, W. Barreto and H. P. de Oliveira, \textit{A spectral domain decomposition code for the Bondi problem}, in preparation.





	






















\end{thebibliography}
\end{document}